\documentclass[aps,prl,twocolumn,notitlepage,superscriptaddress]{revtex4-1}
\usepackage[colorlinks=true, citecolor=magenta, linkcolor=blue, urlcolor=blue]{hyperref}
\usepackage{graphicx}
\usepackage{amsmath}
\usepackage{amssymb}
\usepackage{amsfonts}
\usepackage{xcolor}
\usepackage{tikz}
\usepackage{soul}

\DeclareMathOperator{\im}{Im}

\newcommand{\bk}{{\bf k}}
\newcommand{\bq}{{\bf q}}

\allowdisplaybreaks

\begin{document}
 
\title{
All-Optical Generation of Antiferromagnetic Magnon Currents via the Magnon Circular Photogalvanic Effect}
\author{Emil Vi\~nas Bostr\"om}
\email{emil.bostrom@mpsd.mpg.de}
\affiliation{Max Planck Institute for the Structure and Dynamics of Matter, Center for Free Electron Laser Science (CFEL), Luruper Chaussee 149, 22761 Hamburg, Germany}
\author{Tahereh Sadat Parvini}
\affiliation{Max Planck Institute for the Science of Light, Staudtstrasse 2, PLZ 91058 Erlangen, Germany}
\affiliation{Institute of Physics, University of Greifswald, Felix-Hausdorff-Str. 6, Greifswald, 17489, Germany}
\author{James W. McIver}
\affiliation{Max Planck Institute for the Structure and Dynamics of Matter, Center for Free Electron Laser Science (CFEL), Luruper Chaussee 149, 22761 Hamburg, Germany}
\author{Angel Rubio}
\affiliation{Max Planck Institute for the Structure and Dynamics of Matter, Center for Free Electron Laser Science (CFEL), Luruper Chaussee 149, 22761 Hamburg, Germany}
\affiliation{Center for Computational Quantum Physics, The Flatiron Institute, 162 Fifth Avenue, New York, NY 10010, United States of America}
\author{Silvia Viola Kusminskiy}
\affiliation{Max Planck Institute for the Science of Light, Staudtstrasse 2, PLZ 91058 Erlangen, Germany}
\affiliation{Institute for Theoretical Physics, University of Erlangen-N\"urnberg, Staudtstrasse 7, 91058 Erlangen, Germany}
\author{Michael A.~Sentef}
\email{michael.sentef@mpsd.mpg.de}
\affiliation{Max Planck Institute for the Structure and Dynamics of Matter, Center for Free Electron Laser Science (CFEL), Luruper Chaussee 149, 22761 Hamburg, Germany}
\date{\today}

\begin{abstract}
We introduce the magnon circular photogalvanic effect enabled by stimulated Raman scattering. This provides an all-optical pathway to the generation of directed magnon currents with circularly polarized light in honeycomb antiferromagnetic insulators. The effect is the leading order contribution to magnon photocurrent generation via optical fields. Control of the magnon current by the polarization and angle of incidence of the laser is demonstrated. Experimental detection by sizeable inverse spin Hall voltages in platinum contacts is proposed.
\end{abstract}

\maketitle

The creation and control of spin currents at the nanoscale are key goals in spintronics and magnonics \cite{barman_2021_2021}. The recent synthesis of quasi two-dimensional layered magnetic insulators, such as transition metal phosphorous trichalcogenides MPX$_3$ (M = Ni, Mn, Fe and X = S, Se) and chromium trihalides CrX$_3$ (X = Cl, I, Br) with a band gap of $\sim 1$ eV \cite{gong_discovery_2017,huang_layer-dependent_2017,song_giant_2018,jiang_electric-field_2018,huang_electrical_2018,gibertini_magnetic_2019,chu_linear_2020}, is inspiring new ideas on how to employ such materials in future spintronics devices \cite{RMP_Tserkovnyak_2018,Jungwirth2018}. One of the most promising approaches is to use optical means for spin current generation and control \cite{Nemec2018,Ishizuka19b}. Importantly, this would allow one to adapt concepts from photocurrent generation in electronic systems, in particular the circular photogalvanic effect (CGPE). The CPGE holds great promise for functionality and applications since it allows to selectively generate currents and probe wavefunction quantum geometry only on the surface, as demonstrated in 3D \cite{mciver_control_2012} and 2D topological insulators \cite{xu_electrically_2018} as well as Weyl materials \cite{ma_direct_2017}. On the other hand, the optical control of magnetization is naturally extended into the ultrafast (THz or faster) regime, as has been demonstrated by coherent control of magnetism in pump-probe experiments \cite{kirilyuk_ultrafast_2010,nova_effective_2017,afanasiev2019lightdriven,disa_polarizing_2020,stupakiewicz_ultrafast_2021,Walowski2016,Schlauderer2019,Siegrist2019}. The prospects of combining the expertise from CPGE and ultrafast magnetism research with antiferromagnetic spintronics and opto-spintronics would pave the way towards a new generation of ultrafast opto-spintronics research with functionality and devices on the horizon. However, there are a number of obstacles to overcome. One of them is the theoretical foundation behind such an approach.

\begin{figure}[!t]
  \includegraphics[width=\columnwidth]{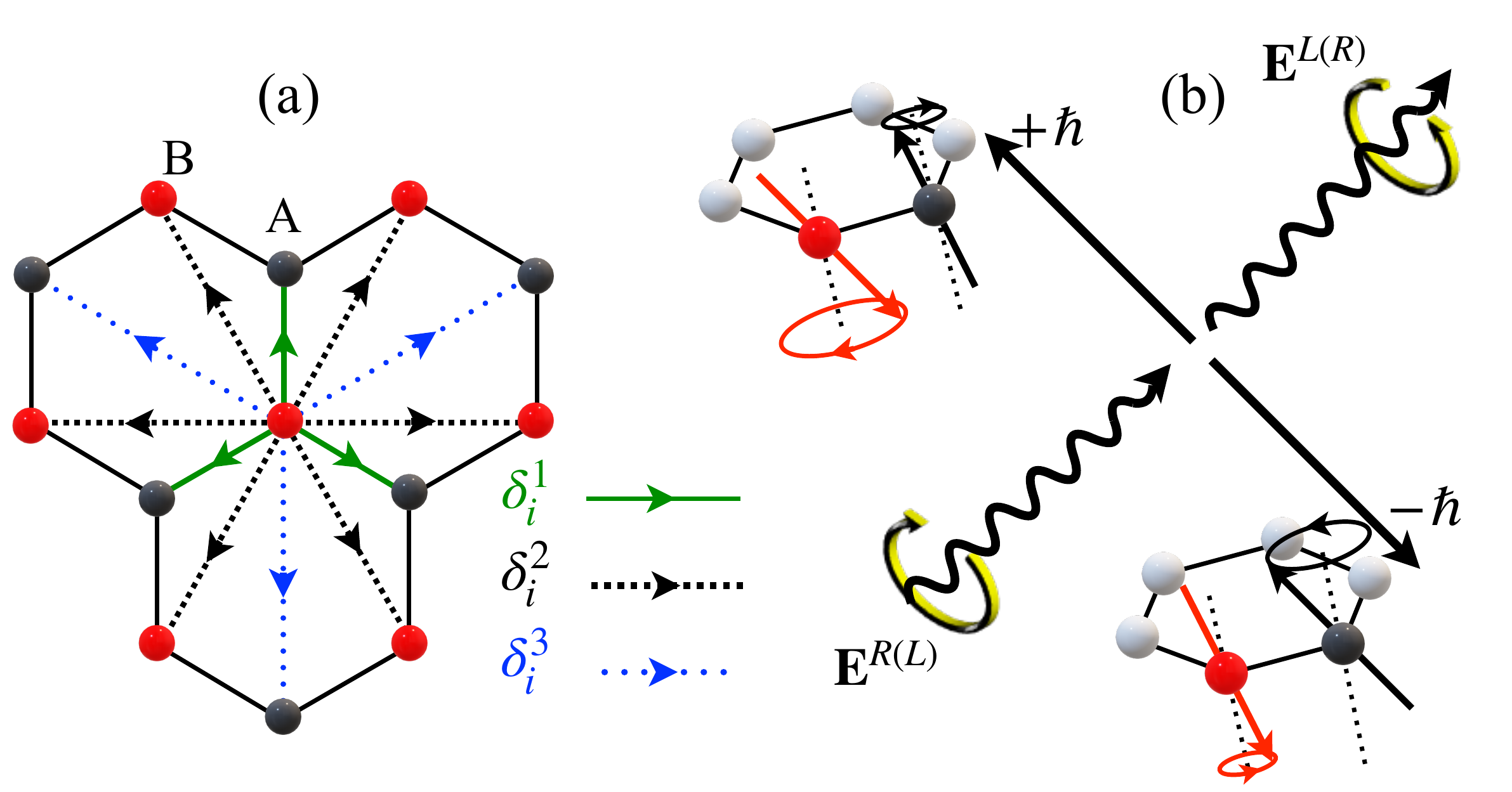}
 \caption{{\bf Magnon photocurrent via stimulated Raman scattering.} $(a)$ The magnetic unit cell with the vectors $\boldsymbol\delta_i^{(n)}$ connecting the $n$th nearest neighbors. $(b)$ Scattering of right-handed (left-handed) photons into left-handed (right-handed) photons imparting spin angular momentum and creating a magnon pair with zero net momentum and spin angular momentum $2\hbar$.}
 \label{fig:setup}
\end{figure}

Here we lay the groundwork for an all-optical route to create directed magnon currents in magnetic insulators through the magnon circular photogalvanic effect (MCPGE) enabled by stimulated Raman scattering. We show that a circularly polarized laser drive generates a magnon current whose strength and direction are controllable through the angle of incidence and polarization via the MCPGE. This magnonic photocurrent is predicted to lead to an inverse spin Hall voltage of experimentally accessible values in platinum contacts, with a characteristic angle dependence, enabling the experimental verification of both the magnon current generation and the underlying MCPGE with existing technology. The MCPGE as proposed in this work is the leading contribution~\cite{Fleury68} to  magnon photocurrents in antiferromagnetic insulators generated by the electric field component of light, in contrast to recent proposals based on the much weaker magnetic dipole interaction~\cite{Proskurin18,Ishizuka19,proskurin2020symmetry}.

\begin{figure}
 \includegraphics[width=0.49\columnwidth]{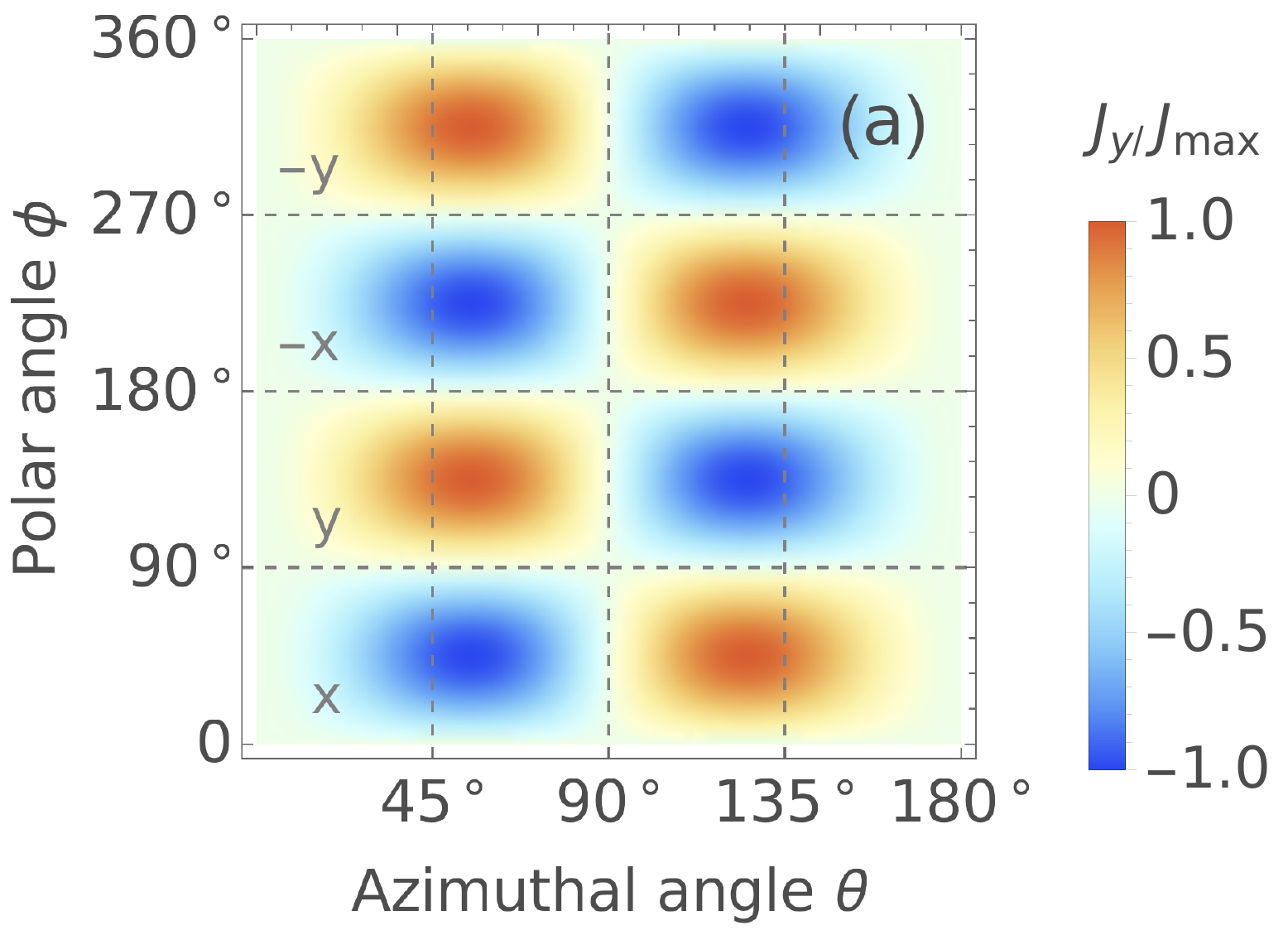}%
 \hfill%
 \includegraphics[width=0.49\columnwidth]{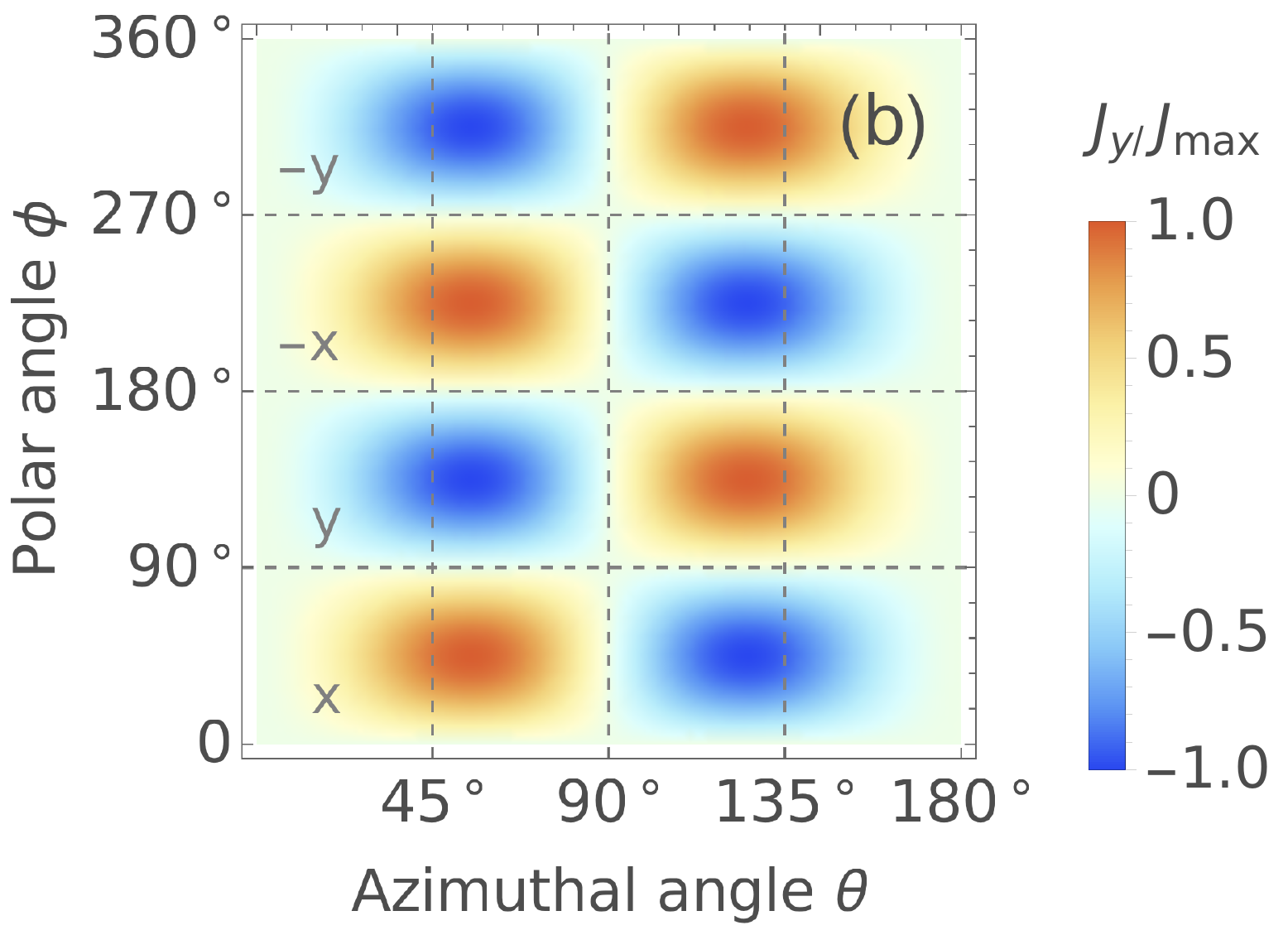}
 \raisebox{-0.5\height}{\includegraphics[width=0.46\columnwidth]{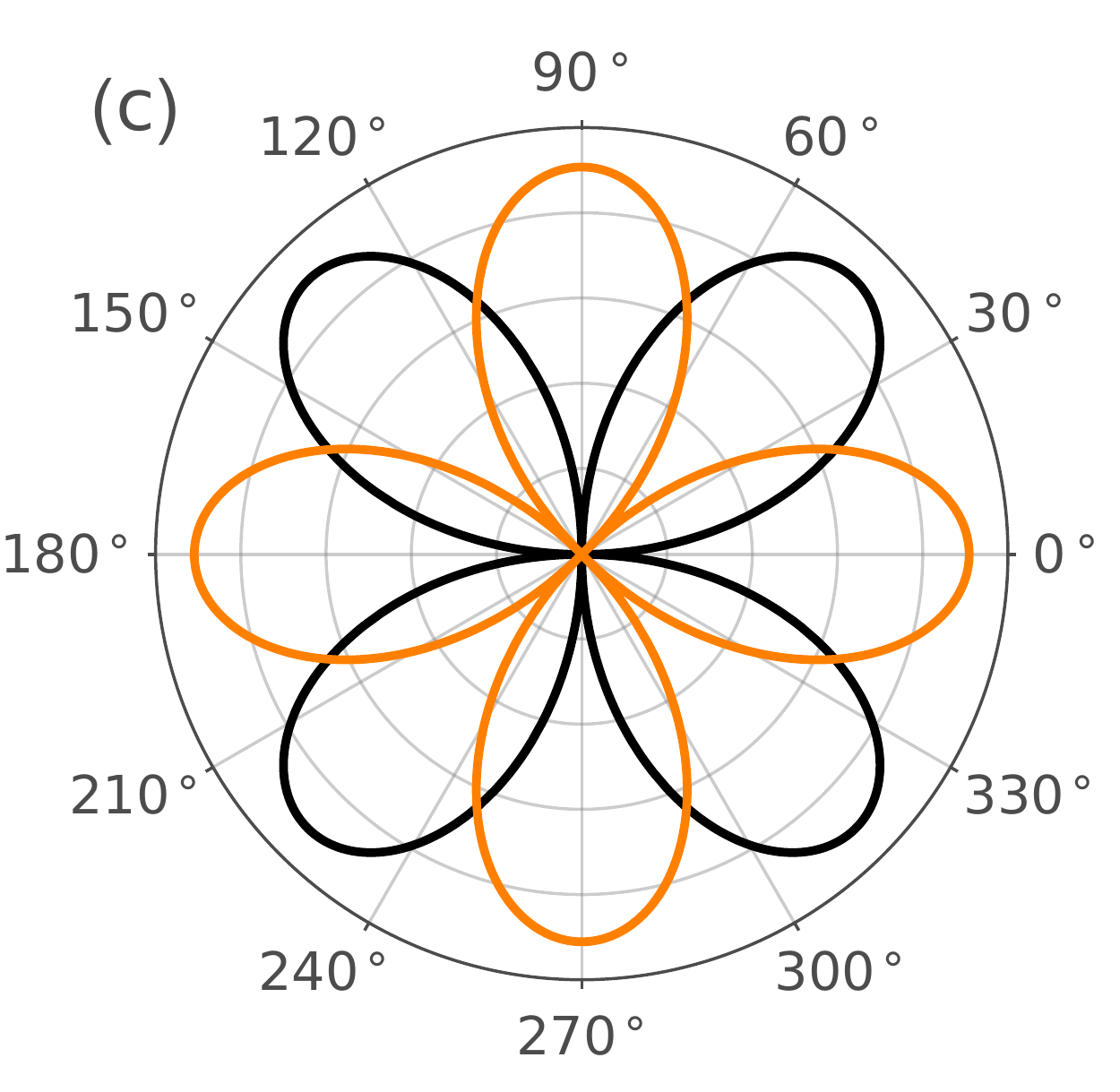}}%
  \hfill%
 \raisebox{-0.5\height}{\includegraphics[width=0.52\columnwidth]{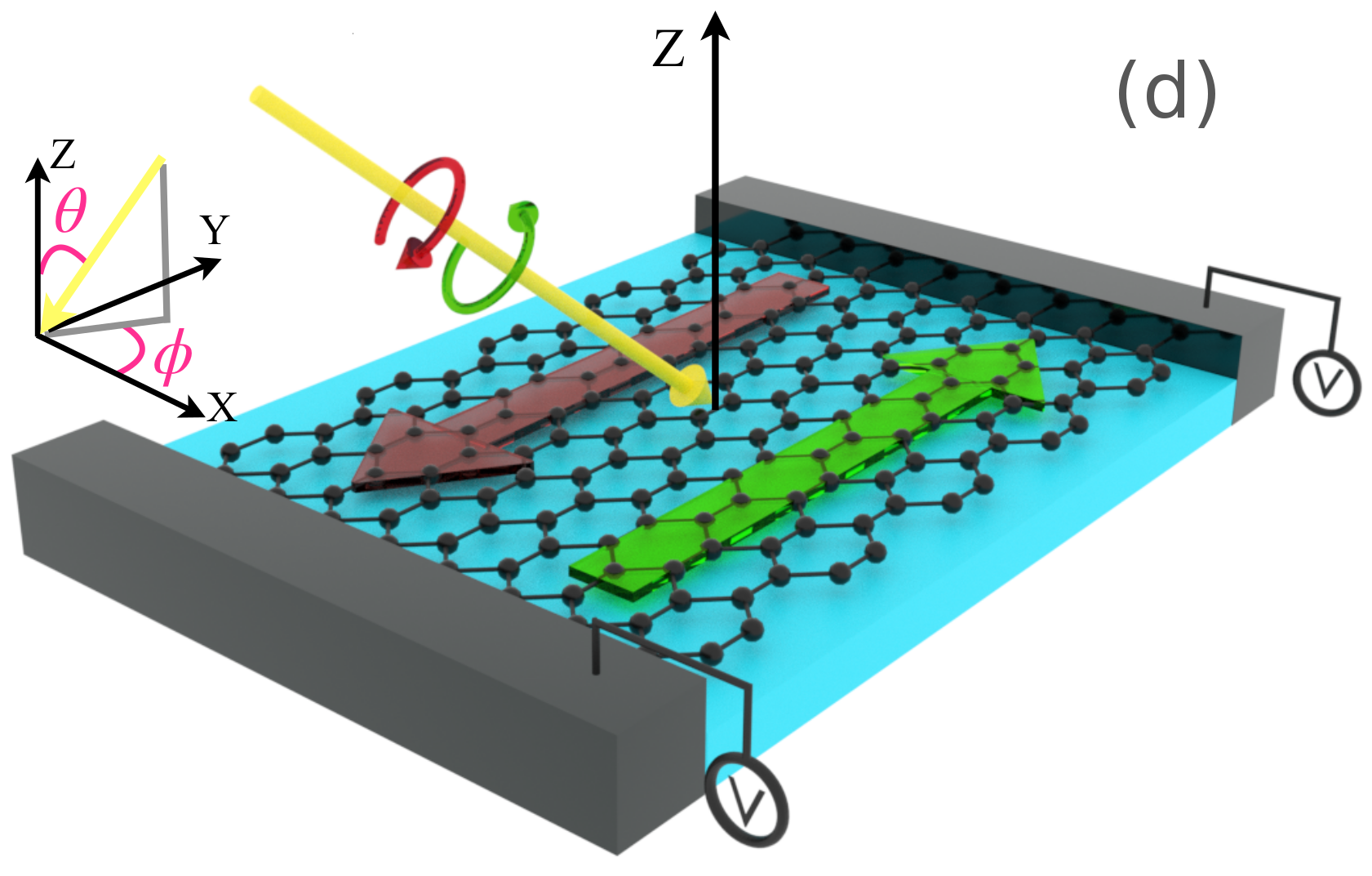}}
 \caption{{\bf Photo-induced magnon current.} $(a,b)$ Photo-induced magnon current along the $y$-axis as a function of incidence and polar angles $\theta$ and $\phi$ for left-handed $(a)$ and right-handed $(b)$ circular polarization. The current is normalized to the maximal value $J_{\rm max}$ obtained at $\theta \approx 55^\circ$ and $\phi = 45^\circ$. The horizontal grid lines indicate where the incident field is parallel to the $x$- or $y$-axis. $(c)$ The $x$-component (orange) and $y$-component (black) of the magnon photocurrent as a function of polar angle $\phi$. $(d)$ Illustration of the proposed experimental setup with a honeycomb antiferromagnet and metallic contacts (Pt) for magnon current read-out. In all panels, the model parameters are $S = 5/2$, $J_1 = 1.54$ meV, $J_2 = -0.14$ meV, $J_3 = 0.3$ meV, $J_z = 8.6$ $\mu$eV and $B = 0$, as appropriate for MnPS$_3$~\cite{Wildes98,Cheng16}.}
 \label{fig:raman}
\end{figure}

In what follows we study the generation of magnon currents via stimulated Raman scattering specifically in collinear honeycomb antiferromagnets. However, the analysis presented below can be straightforwardly extended to general magnetic point groups. We find that the magnon current is determined by the MCPGE and given by  
\begin{align}\label{eq:magnon_current}
 \langle {\bf J} \rangle =  \zeta \im(\sigma) \cos\theta \sin^2 \theta (\sin2\phi \hat{\bf e}_y - \cos2\phi \hat{\bf e}_x),
\end{align}
where $\sigma$ is the non-zero element of the optical susceptibility and $\theta$ and $\phi$ determine the propagation direction of the light as shown in Fig.~\ref{fig:raman}. For subgap excitations the susceptibility only shows a weak frequency dependence and in the following we assume $\hbar\omega = 1$ eV. The magnitude of $\langle {\bf J} \rangle$ is controlled by the angle of incidence $\theta$, while its direction is determined by the polar angle $\phi$ and the chirality $\zeta$ of the laser. The current vanishes at normal and in-plane incidence, and its direction rotates in the substrate plane with a period $2\phi$ as illustrated by the flower shape in Fig.~\ref{fig:raman}$(c)$. Fig.~\ref{fig:raman}$(a,b)$ shows the $y$-component of the current as a function of $\theta$ and $\phi$, for a laser with left- and right-handed polarization, respectively. The results clearly illustrate the direct proportionality of $\langle {\bf J} \rangle$ to the chirality of the laser, and how the current can be controlled via the MCPGE.

The expression for the magnon current $ \langle {\bf J} \rangle $ can be understood from a symmetry analysis of the system. The light-matter coupling due to stimulated Raman scattering is quadratic in the electric fields~\cite{Fleury67,Fleury68,Shastry90}. However, as the scattered photons are not detected the current is given by an integral over scattered photon states. This leaves a quadratic dependence on the incident electric field in line with previous work~\cite{mciver_control_2012,xu_electrically_2018,ma_direct_2017,Fei21}. Since the magnon photo-current is obtained by expanding the optical susceptibility to lowest order in the Raman interaction, only odd orders will contribute. The third order susceptibility vanishes by symmetry, and hence the leading order contribution comes from the fifth order tensor $\sigma_{ijklm}$. For a collinear Ne\'el state the system has $C_{3v}$ symmetry, and the susceptibility has to be invariant under the corresponding symmetry transformations. The $C_{3v}$ and index permutation symmetries reduce the original 32 elements of $\sigma_{ijklm}$ to 16 non-zero elements, out of which three are independent~\cite{SM}. Among these, only one corresponds to a process of net angular momentum transfer that can generate a non-zero magnon current, leading to Eq.~\eqref{eq:magnon_current}.

\begin{figure}
 \raisebox{-0.5\height}{\includegraphics[width=0.5\columnwidth]{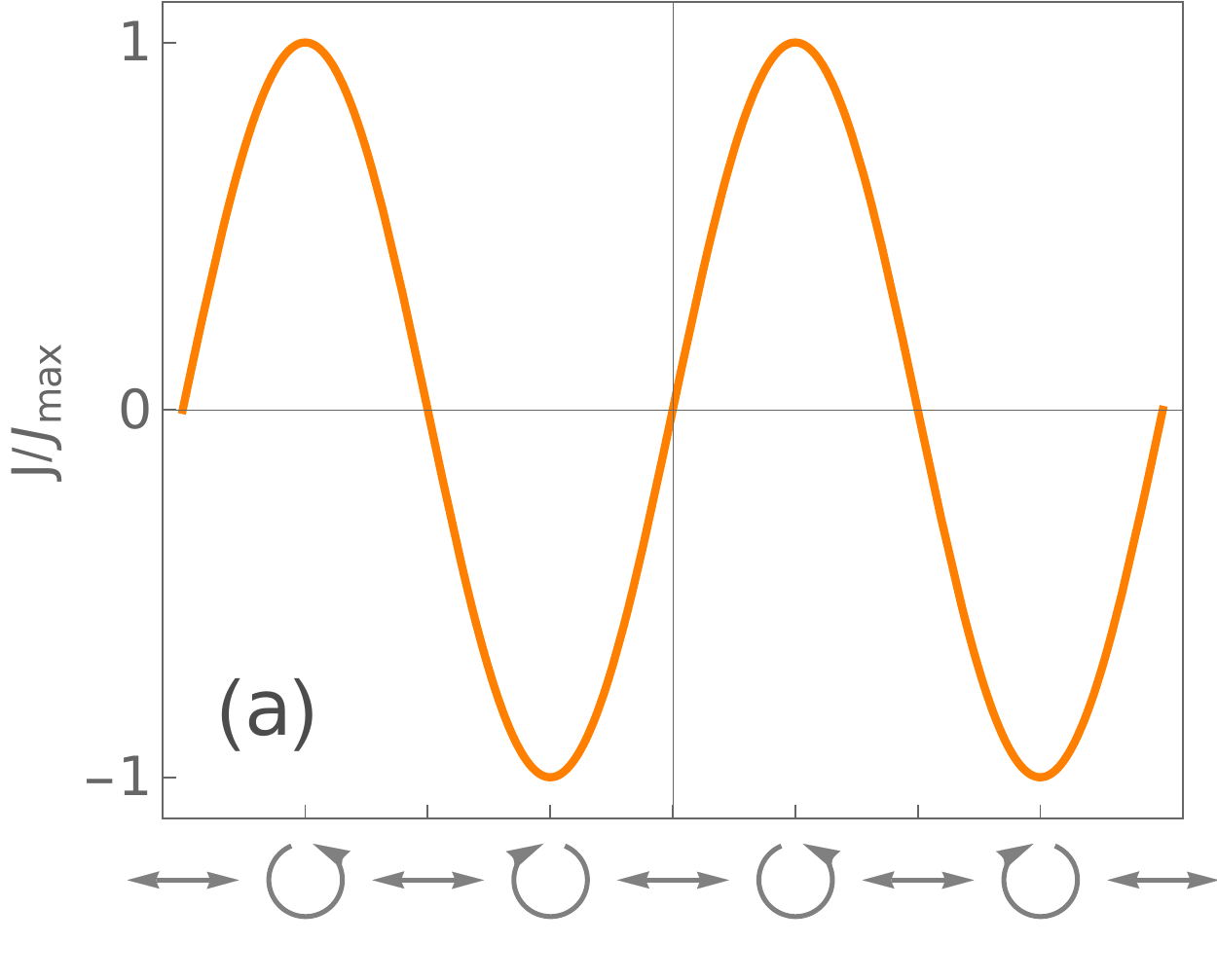}}%
  \hfill%
 \raisebox{-0.5\height}{\includegraphics[width=0.49\columnwidth]{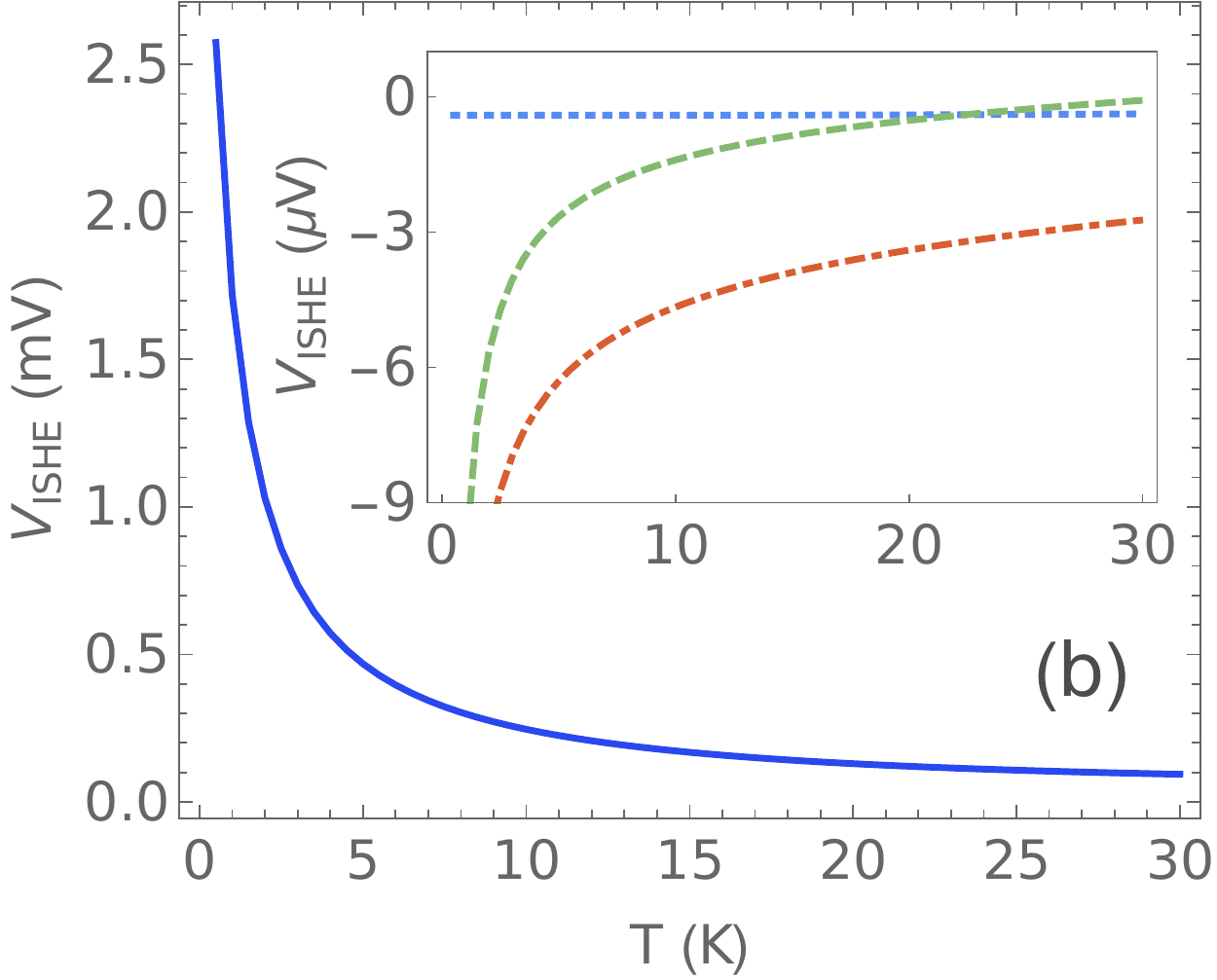}}
 \caption{{\bf Experimental signatures.} $(a)$ Magnon photo-current as a function of polarization obtained by varying the angle between the electric field of a linearly polarized laser and the fast axis of a quarter wave plate. The resulting polarization is indicated below the graph. $(b)$ Inverse spin Hall effect voltage $V_{\rm ISHE}$ induced by a magnon current as a function of temperature $T$. The main panel shows the diagonal Stokes component, while the inset shows the off-diagonal Stokes (blue, dotted), and diagonal (green, dashed) and off-diagonal (brown, dot-dashed) anti-Stokes contributions to $V_{\rm ISHE}$. In both panels, the parameters are the same as in Fig.~\ref{fig:raman}.}
 \label{fig:experiment}
\end{figure}

In a typical experiment the sample and angle of incidence are held fixed while the polarization is varied via a quarter wave plate. Therefore, we show in Fig.~\ref{fig:experiment}$(a)$ the magnon photo-current as a function of polarization for a given configuration $(\theta,\phi)$. The current is maximal for circular polarization, and gradually reverses its direction when the polarization is tuned from left- to right-handed. Incidentally, we find that the current vanishes for linear polarization. The necessity for circularly polarized light can be understood from the requirement of angular momentum conservation, as indicated in Fig.~\ref{fig:setup}.

The magnon current can be detected via the inverse spin Hall effect (ISHE) using the setup proposed in Fig.~\ref{fig:raman}: A magnon current generated in the bulk antiferromagnet propagates towards the Pt contacts, where the resulting magnon accumulation is converted into a spin current. The spin current in turn generates a charge current via the ISHE, which induces a voltage $V_{\rm ISHE}$~\cite{Cornelissen15,Lebrun18}. Following the discussion in Refs.~\cite{Ando11,Zhang12}, we converted the magnon photocurrent into the ISHE voltage shown in Fig.~\ref{fig:experiment}$(b)$~\cite{SM}. To phenomenologically account for the effects of magnon-magnon and magnon-phonon scattering at finite temperatures, a linear temperature dependence of the magnon decay rate is assumed. We find a voltage $V_{\rm ISHE} \sim 1$ mV, of the same order of magnitude as signals detected from DC magnon currents launched via the ISHE~\cite{Wei14}. Fig.~\ref{fig:experiment}$(b)$ shows that the current decays rapidly with temperature, and is effectively zero for $T \approx 30$ K. This is in line with experiments on MnPS$_3$~\cite{Xing19} where the magnon current was found to vanish above $T \approx 30$ K, far below the Ne\'el temperature $T_N \approx 80$ K.

The symmetry analysis of collinear honeycomb antiferromagnets shows the generality of a non-zero magnon current generated by the MCPGE, independent of the specifics of the underlying spin Hamiltonian. This indicates that the mechanism is a generic feature of a large class of materials, which should favor an experimental verification. In what follows we derive Eq.~\eqref{eq:magnon_current} from a concrete microscopic spin Hamiltonian. 

The magnetic structure of a collinear honeycomb antiferromagnet is described by the Hamiltonian
\begin{align}\label{eq:spin_ham}
 H &= \sum_{\langle ij\rangle}J_{ij} {\bf S}_i \cdot {\bf S}_j  + J_z \sum_{\langle ij\rangle} S_i^z S_j^z  - g\mu_B {\bf B}_0 \cdot \sum_i {\bf S}_i,
\end{align}
where $J_{ij}$ are bilinear exchange interactions, $J_z$ is an easy-axis anisotropy, and ${\bf B}_0 = B_0 \hat{\bf z}$ is an external magnetic field along the easy axis. The ground state is approximately given by the Ne\'el state with spins on sublattice $A$ ($B$) pointing in the positive (negative) $z$-direction. This model is appropriate for vdW materials like MnPS$_3$ in the monolayer or weak interlayer coupling limit ~\cite{Wildes, Kim_2019,chittari2016electronic, bazazzadeh2021symmetry}. Eq.~\eqref{eq:spin_ham} can be augmented by including Dzyaloshinskii-Moriya interactions (DMI). Interestingly, we find that the MCPGE is \textit{independent of the DMI}, as discussed further below.

The low-energy excitations of $H$ to lowest order in $1/S$ are found by Holstein-Primakoff linear spin-wave theory. Transforming to Fourier space the Hamiltonian is $H = S \sum_{\bf k} \Psi_{\bf k}^\dagger H_{\bf k} \Psi_{\bf k}$, where $\Psi_{\bf k}^\dagger = (a_{\bf k}^\dagger , b_{\bf -k})$ is a Nambu spinor, $H_{\bf k} = h_0 {\bf 1} + {\bf h} \cdot \boldsymbol\tau$, and $\boldsymbol\tau$ is the vector of Pauli matrices. Including exchange interactions up to third nearest neighbors, the components of the Hamiltonian are given by $h_0 = J + 2J_2  \sum_i \cos({\bf k} \cdot \boldsymbol\delta_i^{(2)})$, 
$h_x - ih_y= \sum_i [J_1 e^{-i{\bf k} \cdot \boldsymbol\delta_i^{(1)}} + J_3 e^{-i {\bf k} \cdot \boldsymbol\delta_i^{(3)}}]$ and $h_z = B/S$, where $J = 3J_1 - 6J_2 +3J_3+ 3J_z$, $B = g\mu_B B_0$ and $\boldsymbol\delta_i^{(n)}$ are the vectors between $n$th nearest neighbors. The Hamiltonian is diagonalized via a paraunitary matrix $U_{\bf k}$ giving $H = \sum_{\bf k} \epsilon_{\alpha\bf k} \alpha_{\bf k}^\dagger \alpha_{\bf k} + \epsilon_{\beta\bf k} \beta_{-\bf k}^\dagger \beta_{-\bf k}$. Here $\epsilon_{\alpha/\beta,\bf k} = d \mp h_z$ is the dispersion of the upper and lower magnon branch respectively, and $d = (h_0^2 - h_x^2 - h_y^2)^{1/2}$. 

We note that the Hamiltonian is invariant under the combined symmetry $\mathcal{TI}$, where $\mathcal{T}$ is the time-reversal operator and $\mathcal{I}$ is a reflection in the inversion center located halfway along an $A - B$ bond. The $\mathcal{TI}$ symmetry implies that the magnon dispersion is even in $\bk$. In the presence of DMIs the $\mathcal{T}\mathcal{I}$ symmetry is broken, and the $C_{6v}$ symmetry of the excitation spectrum reduces to that of the $C_{3v}$ subgroup. Surprisingly, this is the only effect of DMIs in our model, and all results presented here are independent of such interactions~\footnote{This follows from the fact that in honeycomb antiferromagnets, the terms of the Hamiltonian describing the DMI are proportional to the identity matrix.}.


The form of the optomagnetic Hamiltonian can be derived from a tight-binding model by considering a half-filled Mott insulator interacting with a photon field~\cite{Shastry90}, which recovers the results by Fleury and Loudon~\cite{Fleury67,Fleury68}. The electromagnetic field couples to the electronic system via standard Peierls substitution with the vector potential ${\bf A}({\bf r}) = \sum_{\bq s} \gamma_\bq (e^{i\bq\cdot{\bf r}} {\bf e}_{\bq s} a_{\bq s} + e^{-i\bq\cdot{\bf r}} {\bf e}^{*}_{\bq s} a_{\bq s}^\dagger)$, where $\gamma_\bq = (\hbar/2\epsilon_0 \omega_\bq V)^{1/2}$ and ${\bf e}_{\bq s}$ is a polarization vector. For sufficiently weak fields and in the dipole approximation, as controlled by the respective parameters $\lambda = e|{\bf A}|a/\hbar \ll 1$ and $|\bq| \ll 1$, the Peierls phases can be expanded and the effective spin Hamiltonian to lowest order in $t/U$ is given by~\cite{Shastry90}
\begin{align}\label{eq:raman_ham}
 H_R &= S \sum_{\bk q'q} R_{\bf qq'} \Phi_{\bf k}^\dagger \begin{pmatrix} r_{\bk q'q} && t_{\bk q'q} \\ t_{\bk q'q}^{*} && r_{\bk q'q} \end{pmatrix}
 \Phi_{\bf k} a_{q'}^\dagger a_{q}.
\end{align}
Here $R_{\bf qq'} = J_1 (ea/\hbar)^2 \gamma_\bq \gamma_{\bf q'}$, and to simplify the notation we have defined $q \equiv \{\bq,s\}$. 

\begin{figure}
 \raisebox{-0.51\height}{\includegraphics[width=0.48\columnwidth]{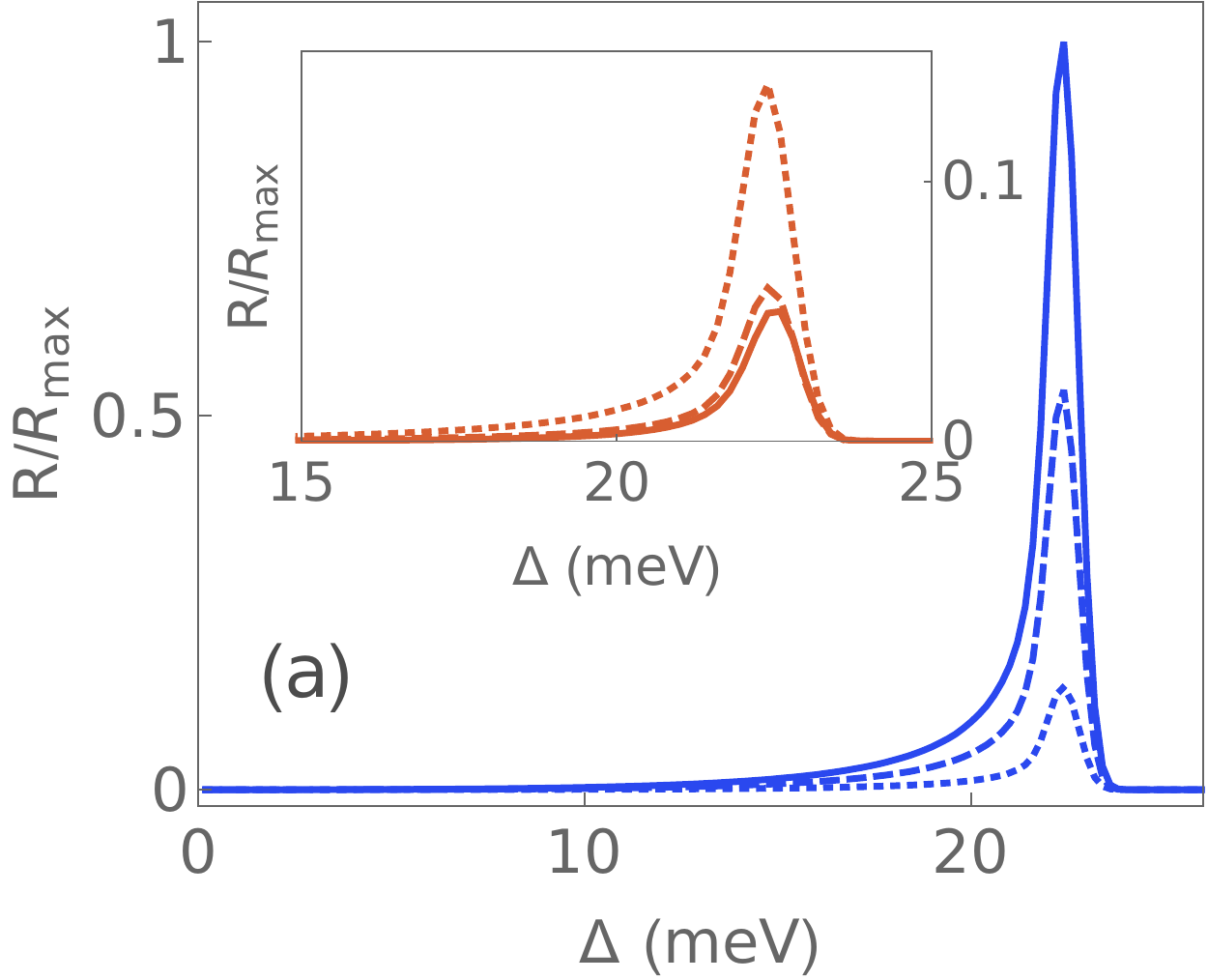}}%
 \hfill%
 \raisebox{-0.5\height}{\includegraphics[width=0.49\columnwidth]{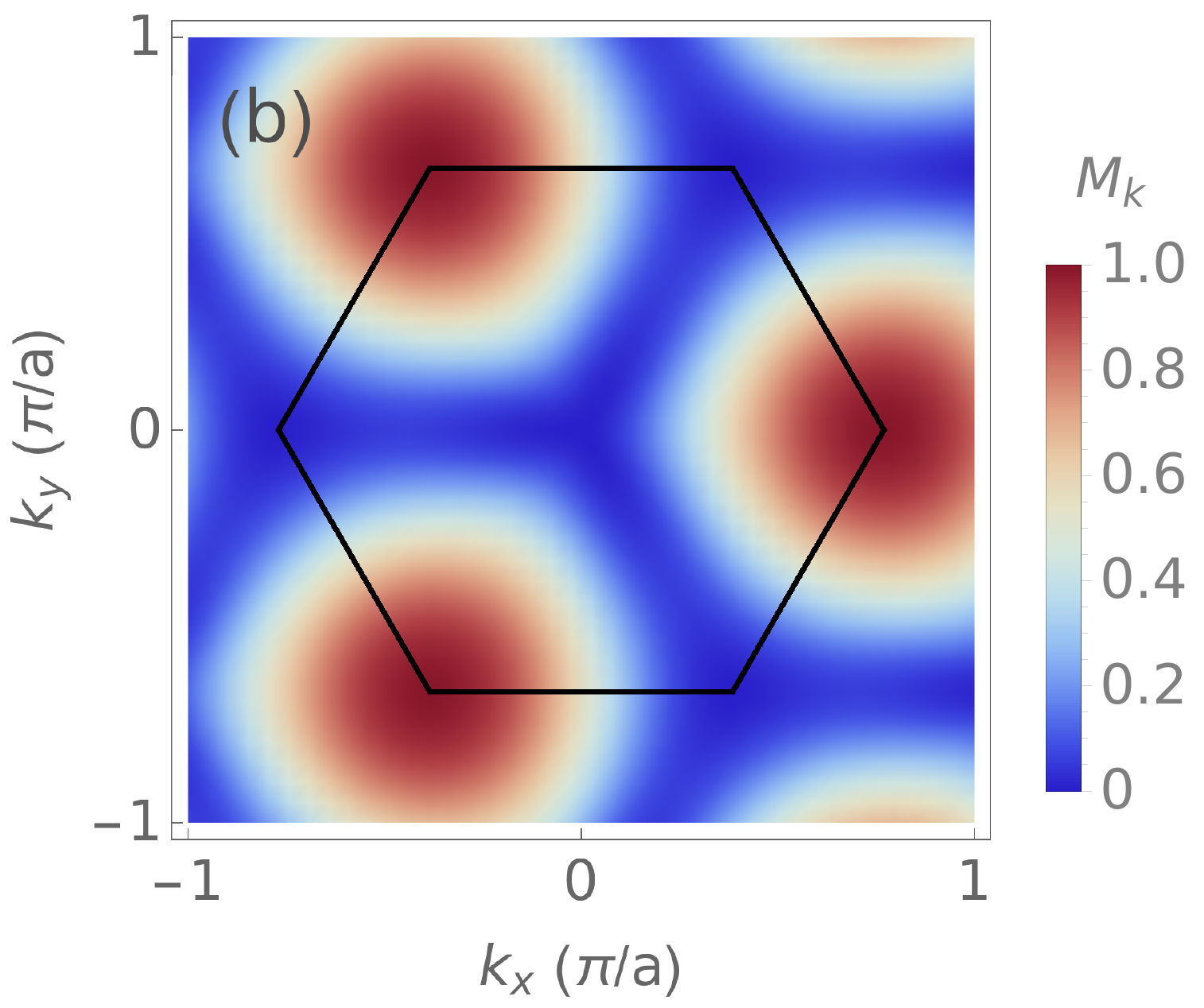}} 
 \caption{{\bf Optical properties of honeycomb antiferromagnets.} $(a)$ Raman cross-section as a function of $\Delta = \hbar\omega_{in} - \hbar\omega_{sc}$ for a right-handed polarized laser at normal incidence (solid lines), and at incidence angles $\theta = \pi/4$ (dashed lines) and $\theta = \pi/2$ (dotted lines). The blue (orange) curves shows the cross-section for scattering into a left-handed (right-handed) photon. $(b)$ Normalized Raman form factor $M_{\bk q_{in}}$ as a function of $\bk$ for a right-handed polarized laser at normal incidence. In all panels, the parameters are the same as in Fig.~\ref{fig:raman}.}
 \label{fig:dispersion}
\end{figure}

The matrix elements $t_{\bk q'q}$ give the form factor $M_{\bk q_{in}} = \sum_q |t_{\bk q_{in}q}|^2$ of the Raman cross-section for two-magnon excitations. In the linear spin wave approximation, the cross-section is given by $R(q_{in}) = \sum_{\bk q} |t_{\bk q_{in}q}|^2 \delta(\epsilon_{\alpha\bk} + \epsilon_{\beta\bk} - \Delta_\bq)$ with $\Delta_\bq = \hbar\omega_{\bq_{in}} - \hbar\omega_\bq$. The cross-section is shown in Fig.~\ref{fig:dispersion}$(a)$ for a right-handed circularly polarized laser, scattered into either a left- or right-handed circularly polarized photon. At normal incidence, there is an almost complete polarization selection favoring scattering into left-handed photons. This is accompanied by an almost perfect selection rule favoring magnon excitation at the $K$ point, as seen by the $\bk$-dependence of the form factor $M_{\bk q_{in}}$ in Fig.~\ref{fig:dispersion}$(b)$. As the angle of incidence is decreased the relative contributions for scattering into left- and right-handed photons approach each other, and become identical for an in-plane laser. This can be understood by considering the projection of the electric field onto the substrate plane: At normal incidence the projected field is circular and angular momentum conservation requires scattering into a mode of opposite handedness. For in-plane incidence the projected field is linear, and thus consists of equal parts left- and right-handed photons.


The optical susceptibility tensor is calculated via the second order Kubo formalism~\cite{Proskurin18,Ishizuka19} using the magnon current operator ${\bf J}$~\cite{SM}. The magnons are assumed to be in initial equilibrium at a temperature $T$ and the photons to be in the initial state $|n_{q_{in}}\rangle \prod_q |0_q\rangle$, with a single macroscopically populated photon mode corresponding to the incident laser. Since the experimental setup of Fig.~\ref{fig:raman} is insensitive to the energy and polarization of the scattered photons, the current is integrated over scattered photon states.

The magnon photocurrent consists of a Stokes and an anti-Stokes component~\cite{SM}, with the latter vanishing in the zero temperature limit. Further, both components have contributions from both the diagonal and off-diagonal terms of the magnon current operator. For the spin parameters considered here~\cite{Wildes98,Cheng16}, the diagonal Stokes term is larger than the remaining contributions by about three orders of magnitude (see Fig.~\ref{fig:experiment}$(d)$), and the photo-induced magnon current is therefore given to a very good approximation by
\begin{align}\label{eq:raman_quadratic}
 \langle {\bf J} \rangle (q_{in}) &= \frac{2G}{\hbar\omega_{\bq_{in}} \Gamma} \sum_{s} \int \frac{d\bk^2}{(2\pi)^2} \, (n_{\bk\alpha} + n_{\bk\beta} + 1) \nonumber \\
 &\hspace*{0.8cm}\times ({\bf v}_{\alpha\bk} + {\bf v}_{\beta\bk}) |t_{\bk qq_{in}}|^2\Delta_\bk.
\end{align}
Here $G$ is a constant depending on the intensity $I$ and photon energy $\hbar\omega_{\bq_{in}}$ of the incident laser, $n_{\bk\alpha}$ ($n_{\bk\beta}$) is the thermal magnon population, $\Gamma^{-1}$ is a phenomenological magnon lifetime, and $\Delta_\bk = \hbar\omega_{\bq_{in}} - \epsilon_{\alpha\bk} - \epsilon_{\beta\bk}$. Taking $I = 10^{12}$ W/cm$^2$ and $\hbar\omega_{\bq_{in}} = 1$ eV, we find $G \approx 3.6 \cdot 10^{-12}$ meV/\AA. 

Since the magnon pairs $|\alpha_\bk,\beta_{-\bk}\rangle$ carry zero net momentum, momentum conservation forces the wave vectors of the incident and scattered photons to be identical. Taking $\bq_{in} = \bq = q(\cos\phi\sin\theta,\sin\phi\sin\theta,-\cos\theta)$, where $\theta$ and $\phi$ are defined in Fig.~\ref{fig:setup}, the polarization of the incident laser can be written as $\hat{\bf e}_{in} = (\hat{\bf e}_1 - i\zeta \hat{\bf e}_2)/\sqrt{2}$. Here $\hat{\bf e}_1$, $\hat{\bf e}_2$ and $\bq$ constitute a right-handed coordinate system and $\zeta = 1$ ($\zeta = -1$) for right-handed (left-handed) polarization. Evaluating the symmetry allowed elements of the susceptibility tensor from Eq.~\ref{eq:raman_quadratic}, only $\sigma = \sigma_{yyxyx}$ is found to give a non-zero contribution to $\langle {\bf J} \rangle$, see Eq.~\eqref{eq:magnon_current}.


To understand the origin and angular dependence of the magnon photocurrent we consider the momentum space structure of the excited magnon population for different $(\theta,\phi)$. At normal incidence ($\theta = 0$) the excited magnon population has $C_3$ rotational symmetry, and depending on the field chirality magnons are excited mainly at $K$ or at $K'$ (see Fig.~\ref{fig:dispersion}$(d)$). Expanding the Raman form factor $M_\bk$ around $K_\tau =  4 \pi\tau/(3 \sqrt{3} a) \hat{\bf e}_x$, with $\tau = \pm 1$ at $K$ ($K'$), we find $M_\bk = 3(RSa)^2/2 (1 - \eta\tau) + \mathcal{O}(k^2)$. Thus, we find a near perfect optical selection rule for two-magnon excitations in honeycomb antiferromagnets. However, the $C_3$ symmetry leads to a vanishing current when integrated over the Brillouin zone. Analogously, at in-plane incidence ($\theta = \pi/2$) the excited magnon population is $C_2$ symmetric, again leading to a vanishing integrated current. In the intermediate range $0 < \theta < \pi/2$ the magnon population is asymmetric and interpolates between the limiting cases at $\theta = 0$ and $\theta = \pi/2$, leading to a maximal current at $\theta = \arctan\sqrt{2} \approx \pi/3$ (see Fig.~\ref{fig:raman}). 

The angular dependence of $\langle {\bf J} \rangle$ is understood by noting that as a function of $\phi$, the maxima of the magnon population rotate around the $K$ or $K'$ points at a distance $d(\theta)$ with $d(0) = 0$ and $d(\pi/2) = a/2$. The period of the rotation results from the coincidence of the in-plane projected polarization at $\phi$ and $\phi + \pi$. This leads to a clockwise (anticlockwise) rotation of the current with a period $\pi$ for right-handed (left-handed) fields.


In conclusion, the current generation process creates magnon pairs $|\alpha_\bk, \beta_{-\bk}\rangle$ with zero net momentum and net angular momentum $2\hbar$. This requires absorption and emission of photons with net angular momentum, and thus circularly polarized light. Since the process is symmetric under the transformation $\bk \to -\bk$, the magnon current and MCPGE requires an asymmetric excited magnon population that is achieved by irradiating the system at an oblique angle. Chiral magnon photo-currents thus arise from the interplay of momentum imbalance induced by the chiral optical field and the net angular momentum carried by magnon pairs.

To summarize, we have proposed an all-optical mechanism to generate magnon photo-currents via two-magnon stimulated Raman scattering. The current is directly proportional to the chirality of the incident laser, and can therefore be controlled via the MCPGE. For realistic fields the magnon current should be measurable with existing technology via the induced inverse spin Hall voltage in a Pt contact. Our results are independent of the details of the microscopic spin Hamiltonian and can be derived solely from the symmetries of the magnetic ground state and the requirement of angular momentum conservation, indicating that the MCPGE should be a generic feature of a large class of antiferromagnetic insulators.

The present symmetry analysis of honeycomb lattice antiferromagnets is straightforwardly extended to general magnetic point groups, and in general we expect the MCPGE current to compete with currents insensitive to the polarization. However, for certain point groups the optical susceptibility vanishes by symmetry, thus prohibiting the generation of magnon photo-currents. This is exemplified by the square lattice antiferromagnet, where an additional site inversion symmetry forces both the photo-current and the magnon Berry curvature to vanish.

Our results suggest a deeper link between the magnon photo-current found in this work and the Berry curvature, which should be explored in future works. We also note the close analogy of the optical selection rules discussed above to the selection rules for interband transitions in electronic honeycomb systems~\cite{Yao08}. The key role of quantum geometry for light-matter interaction has recently been noticed in the contexts of electronic flat-band systems~\cite{Topp21} and nonlinear optical responses~\cite{li_detection_2020,ahn_low-frequency_2020,ahn_riemannian_2021}, and one can expect a similar role for the quantum geometry of magnon wavefunctions on the photon-magnon interaction. A natural extension of our work is to consider the photo-induced magnon current in topological magnetic systems where the integrated Berry curvature is non-zero~\cite{Bostrom20}, and where a quantized response could be present ~\cite{de_juan_quantized_2017,ni_giant_2021}.


\begin{acknowledgments}
This  work  was  supported  by  the  European  Research  Council (ERC-2015-AdG694097), the Cluster of Excellence ’Advanced  Imaging  of  Matter’  (AIM),  Grupos  Consolidados (IT1249-19) and SFB925 ”Light  induced  dynamics and control of correlated quantum systems”. The Flatiron Institute is a division of the Simons Foundation. MAS acknowledges financial support through the Deutsche Forschungsgemeinschaft (DFG, German Research Foundation) via the Emmy Noether program (SE 2558/2). AR, JWM and MAS acknowledge support from the Max Planck-New York City Center for Non-Equilibrium Quantum Phenomena. TSP and SVK acknowledge financial support from the Max Planck Society through a Max Planck Research Group. JWM acknowledges support from the Cluster of Excellence ‘CUI: Advanced Imaging of Matter’ of the Deutsche Forschungsgemeinschaft (DFG), EXC 2056, project ID 390715994 and is funded by the Deutsche Forschungsgemeinschaft (DFG, German Research Foundation) – SFB-925 – project 170620586.
\end{acknowledgments}

\bibliography{references_circular,Magnon_Transport}

\begin{thebibliography}{52}%
\makeatletter
\providecommand \@ifxundefined [1]{%
 \@ifx{#1\undefined}
}%
\providecommand \@ifnum [1]{%
 \ifnum #1\expandafter \@firstoftwo
 \else \expandafter \@secondoftwo
 \fi
}%
\providecommand \@ifx [1]{%
 \ifx #1\expandafter \@firstoftwo
 \else \expandafter \@secondoftwo
 \fi
}%
\providecommand \natexlab [1]{#1}%
\providecommand \enquote  [1]{``#1''}%
\providecommand \bibnamefont  [1]{#1}%
\providecommand \bibfnamefont [1]{#1}%
\providecommand \citenamefont [1]{#1}%
\providecommand \href@noop [0]{\@secondoftwo}%
\providecommand \href [0]{\begingroup \@sanitize@url \@href}%
\providecommand \@href[1]{\@@startlink{#1}\@@href}%
\providecommand \@@href[1]{\endgroup#1\@@endlink}%
\providecommand \@sanitize@url [0]{\catcode `\\12\catcode `\$12\catcode
  `\&12\catcode `\#12\catcode `\^12\catcode `\_12\catcode `\%12\relax}%
\providecommand \@@startlink[1]{}%
\providecommand \@@endlink[0]{}%
\providecommand \url  [0]{\begingroup\@sanitize@url \@url }%
\providecommand \@url [1]{\endgroup\@href {#1}{\urlprefix }}%
\providecommand \urlprefix  [0]{URL }%
\providecommand \Eprint [0]{\href }%
\providecommand \doibase [0]{http://dx.doi.org/}%
\providecommand \selectlanguage [0]{\@gobble}%
\providecommand \bibinfo  [0]{\@secondoftwo}%
\providecommand \bibfield  [0]{\@secondoftwo}%
\providecommand \translation [1]{[#1]}%
\providecommand \BibitemOpen [0]{}%
\providecommand \bibitemStop [0]{}%
\providecommand \bibitemNoStop [0]{.\EOS\space}%
\providecommand \EOS [0]{\spacefactor3000\relax}%
\providecommand \BibitemShut  [1]{\csname bibitem#1\endcsname}%
\let\auto@bib@innerbib\@empty
\bibitem [{\citenamefont {Barman}\ \emph {et~al.}(2021)\citenamefont {Barman},
  \citenamefont {Gubbiotti}, \citenamefont {Ladak}, \citenamefont {Adeyeye},
  \citenamefont {Krawczyk}, \citenamefont {Gr{\"a}fe}, \citenamefont
  {Adelmann}, \citenamefont {Cotofana}, \citenamefont {Naeemi}, \citenamefont
  {Vasyuchka}, \citenamefont {Hillebrands}, \citenamefont {Nikitov},
  \citenamefont {Yu}, \citenamefont {Grundler}, \citenamefont {Sadovnikov},
  \citenamefont {Grachev}, \citenamefont {Sheshukova}, \citenamefont
  {Duquesne}, \citenamefont {Marangolo}, \citenamefont {Gyorgy}, \citenamefont
  {Porod}, \citenamefont {Demidov}, \citenamefont {Urazhdin}, \citenamefont
  {Demokritov}, \citenamefont {Albisetti}, \citenamefont {Petti}, \citenamefont
  {Bertacco}, \citenamefont {Schulteiss}, \citenamefont {Kruglyak},
  \citenamefont {Poimanov}, \citenamefont {Sahoo}, \citenamefont {Sinha},
  \citenamefont {Yang}, \citenamefont {Muenzenberg}, \citenamefont {Moriyama},
  \citenamefont {Mizukami}, \citenamefont {Landeros}, \citenamefont {Gallardo},
  \citenamefont {Carlotti}, \citenamefont {Kim}, \citenamefont {Stamps},
  \citenamefont {Camley}, \citenamefont {Rana}, \citenamefont {Otani},
  \citenamefont {Yu}, \citenamefont {Yu}, \citenamefont {Bauer}, \citenamefont
  {Back}, \citenamefont {Uhrig}, \citenamefont {Dobrovolskiy}, \citenamefont
  {Dijken}, \citenamefont {Budinska}, \citenamefont {Qin}, \citenamefont
  {Chumak}, \citenamefont {Khitun}, \citenamefont {Nikonov}, \citenamefont
  {Young}, \citenamefont {Zingsem},\ and\ \citenamefont
  {Winklhofer}}]{barman_2021_2021}%
  \BibitemOpen
  \bibfield  {author} {\bibinfo {author} {\bibfnamefont {A.}~\bibnamefont
  {Barman}}, \bibinfo {author} {\bibfnamefont {G.}~\bibnamefont {Gubbiotti}},
  \bibinfo {author} {\bibfnamefont {S.}~\bibnamefont {Ladak}}, \bibinfo
  {author} {\bibfnamefont {A.~O.}\ \bibnamefont {Adeyeye}}, \bibinfo {author}
  {\bibfnamefont {M.}~\bibnamefont {Krawczyk}}, \bibinfo {author}
  {\bibfnamefont {J.}~\bibnamefont {Gr{\"a}fe}}, \bibinfo {author}
  {\bibfnamefont {C.}~\bibnamefont {Adelmann}}, \bibinfo {author}
  {\bibfnamefont {S.}~\bibnamefont {Cotofana}}, \bibinfo {author}
  {\bibfnamefont {A.}~\bibnamefont {Naeemi}}, \bibinfo {author} {\bibfnamefont
  {V.~I.}\ \bibnamefont {Vasyuchka}}, \bibinfo {author} {\bibfnamefont
  {B.}~\bibnamefont {Hillebrands}}, \bibinfo {author} {\bibfnamefont {S.~A.}\
  \bibnamefont {Nikitov}}, \bibinfo {author} {\bibfnamefont {H.}~\bibnamefont
  {Yu}}, \bibinfo {author} {\bibfnamefont {D.}~\bibnamefont {Grundler}},
  \bibinfo {author} {\bibfnamefont {A.}~\bibnamefont {Sadovnikov}}, \bibinfo
  {author} {\bibfnamefont {A.~A.}\ \bibnamefont {Grachev}}, \bibinfo {author}
  {\bibfnamefont {S.~E.}\ \bibnamefont {Sheshukova}}, \bibinfo {author}
  {\bibfnamefont {J.-Y.}\ \bibnamefont {Duquesne}}, \bibinfo {author}
  {\bibfnamefont {M.}~\bibnamefont {Marangolo}}, \bibinfo {author}
  {\bibfnamefont {C.}~\bibnamefont {Gyorgy}}, \bibinfo {author} {\bibfnamefont
  {W.}~\bibnamefont {Porod}}, \bibinfo {author} {\bibfnamefont {V.~E.}\
  \bibnamefont {Demidov}}, \bibinfo {author} {\bibfnamefont {S.}~\bibnamefont
  {Urazhdin}}, \bibinfo {author} {\bibfnamefont {S.}~\bibnamefont
  {Demokritov}}, \bibinfo {author} {\bibfnamefont {E.}~\bibnamefont
  {Albisetti}}, \bibinfo {author} {\bibfnamefont {D.}~\bibnamefont {Petti}},
  \bibinfo {author} {\bibfnamefont {R.}~\bibnamefont {Bertacco}}, \bibinfo
  {author} {\bibfnamefont {H.}~\bibnamefont {Schulteiss}}, \bibinfo {author}
  {\bibfnamefont {V.~V.}\ \bibnamefont {Kruglyak}}, \bibinfo {author}
  {\bibfnamefont {V.~D.}\ \bibnamefont {Poimanov}}, \bibinfo {author}
  {\bibfnamefont {A.~K.}\ \bibnamefont {Sahoo}}, \bibinfo {author}
  {\bibfnamefont {J.}~\bibnamefont {Sinha}}, \bibinfo {author} {\bibfnamefont
  {H.}~\bibnamefont {Yang}}, \bibinfo {author} {\bibfnamefont {M.}~\bibnamefont
  {Muenzenberg}}, \bibinfo {author} {\bibfnamefont {T.}~\bibnamefont
  {Moriyama}}, \bibinfo {author} {\bibfnamefont {S.}~\bibnamefont {Mizukami}},
  \bibinfo {author} {\bibfnamefont {P.}~\bibnamefont {Landeros}}, \bibinfo
  {author} {\bibfnamefont {R.~A.}\ \bibnamefont {Gallardo}}, \bibinfo {author}
  {\bibfnamefont {G.}~\bibnamefont {Carlotti}}, \bibinfo {author}
  {\bibfnamefont {J.-V.}\ \bibnamefont {Kim}}, \bibinfo {author} {\bibfnamefont
  {R.~L.}\ \bibnamefont {Stamps}}, \bibinfo {author} {\bibfnamefont {R.~E.}\
  \bibnamefont {Camley}}, \bibinfo {author} {\bibfnamefont {B.}~\bibnamefont
  {Rana}}, \bibinfo {author} {\bibfnamefont {Y.}~\bibnamefont {Otani}},
  \bibinfo {author} {\bibfnamefont {W.}~\bibnamefont {Yu}}, \bibinfo {author}
  {\bibfnamefont {T.}~\bibnamefont {Yu}}, \bibinfo {author} {\bibfnamefont
  {G.~E.~W.}\ \bibnamefont {Bauer}}, \bibinfo {author} {\bibfnamefont {C.~H.}\
  \bibnamefont {Back}}, \bibinfo {author} {\bibfnamefont {G.~S.}\ \bibnamefont
  {Uhrig}}, \bibinfo {author} {\bibfnamefont {O.~V.}\ \bibnamefont
  {Dobrovolskiy}}, \bibinfo {author} {\bibfnamefont {S.~v.}\ \bibnamefont
  {Dijken}}, \bibinfo {author} {\bibfnamefont {B.}~\bibnamefont {Budinska}},
  \bibinfo {author} {\bibfnamefont {H.}~\bibnamefont {Qin}}, \bibinfo {author}
  {\bibfnamefont {A.}~\bibnamefont {Chumak}}, \bibinfo {author} {\bibfnamefont
  {A.}~\bibnamefont {Khitun}}, \bibinfo {author} {\bibfnamefont {D.~E.}\
  \bibnamefont {Nikonov}}, \bibinfo {author} {\bibfnamefont {I.~A.}\
  \bibnamefont {Young}}, \bibinfo {author} {\bibfnamefont {B.}~\bibnamefont
  {Zingsem}}, \ and\ \bibinfo {author} {\bibfnamefont {M.}~\bibnamefont
  {Winklhofer}},\ }\href {\doibase 10.1088/1361-648X/abec1a} {\bibfield
  {journal} {\bibinfo  {journal} {Journal of Physics: Condensed Matter}\ }
  (\bibinfo {year} {2021}),\ 10.1088/1361-648X/abec1a}\BibitemShut {NoStop}%
\bibitem [{\citenamefont {Gong}\ \emph {et~al.}(2017)\citenamefont {Gong},
  \citenamefont {Li}, \citenamefont {Li}, \citenamefont {Ji}, \citenamefont
  {Stern}, \citenamefont {Xia}, \citenamefont {Cao}, \citenamefont {Bao},
  \citenamefont {Wang}, \citenamefont {Wang}, \citenamefont {Qiu},
  \citenamefont {Cava}, \citenamefont {Louie}, \citenamefont {Xia},\ and\
  \citenamefont {Zhang}}]{gong_discovery_2017}%
  \BibitemOpen
  \bibfield  {author} {\bibinfo {author} {\bibfnamefont {C.}~\bibnamefont
  {Gong}}, \bibinfo {author} {\bibfnamefont {L.}~\bibnamefont {Li}}, \bibinfo
  {author} {\bibfnamefont {Z.}~\bibnamefont {Li}}, \bibinfo {author}
  {\bibfnamefont {H.}~\bibnamefont {Ji}}, \bibinfo {author} {\bibfnamefont
  {A.}~\bibnamefont {Stern}}, \bibinfo {author} {\bibfnamefont
  {Y.}~\bibnamefont {Xia}}, \bibinfo {author} {\bibfnamefont {T.}~\bibnamefont
  {Cao}}, \bibinfo {author} {\bibfnamefont {W.}~\bibnamefont {Bao}}, \bibinfo
  {author} {\bibfnamefont {C.}~\bibnamefont {Wang}}, \bibinfo {author}
  {\bibfnamefont {Y.}~\bibnamefont {Wang}}, \bibinfo {author} {\bibfnamefont
  {Z.~Q.}\ \bibnamefont {Qiu}}, \bibinfo {author} {\bibfnamefont {R.~J.}\
  \bibnamefont {Cava}}, \bibinfo {author} {\bibfnamefont {S.~G.}\ \bibnamefont
  {Louie}}, \bibinfo {author} {\bibfnamefont {J.}~\bibnamefont {Xia}}, \ and\
  \bibinfo {author} {\bibfnamefont {X.}~\bibnamefont {Zhang}},\ }\href
  {\doibase 10.1038/nature22060} {\bibfield  {journal} {\bibinfo  {journal}
  {Nature}\ }\textbf {\bibinfo {volume} {546}},\ \bibinfo {pages} {265}
  (\bibinfo {year} {2017})}\BibitemShut {NoStop}%
\bibitem [{\citenamefont {Huang}\ \emph {et~al.}(2017)\citenamefont {Huang},
  \citenamefont {Clark}, \citenamefont {Navarro-Moratalla}, \citenamefont
  {Klein}, \citenamefont {Cheng}, \citenamefont {Seyler}, \citenamefont
  {Zhong}, \citenamefont {Schmidgall}, \citenamefont {McGuire}, \citenamefont
  {Cobden}, \citenamefont {Yao}, \citenamefont {Xiao}, \citenamefont
  {Jarillo-Herrero},\ and\ \citenamefont {Xu}}]{huang_layer-dependent_2017}%
  \BibitemOpen
  \bibfield  {author} {\bibinfo {author} {\bibfnamefont {B.}~\bibnamefont
  {Huang}}, \bibinfo {author} {\bibfnamefont {G.}~\bibnamefont {Clark}},
  \bibinfo {author} {\bibfnamefont {E.}~\bibnamefont {Navarro-Moratalla}},
  \bibinfo {author} {\bibfnamefont {D.~R.}\ \bibnamefont {Klein}}, \bibinfo
  {author} {\bibfnamefont {R.}~\bibnamefont {Cheng}}, \bibinfo {author}
  {\bibfnamefont {K.~L.}\ \bibnamefont {Seyler}}, \bibinfo {author}
  {\bibfnamefont {D.}~\bibnamefont {Zhong}}, \bibinfo {author} {\bibfnamefont
  {E.}~\bibnamefont {Schmidgall}}, \bibinfo {author} {\bibfnamefont {M.~A.}\
  \bibnamefont {McGuire}}, \bibinfo {author} {\bibfnamefont {D.~H.}\
  \bibnamefont {Cobden}}, \bibinfo {author} {\bibfnamefont {W.}~\bibnamefont
  {Yao}}, \bibinfo {author} {\bibfnamefont {D.}~\bibnamefont {Xiao}}, \bibinfo
  {author} {\bibfnamefont {P.}~\bibnamefont {Jarillo-Herrero}}, \ and\ \bibinfo
  {author} {\bibfnamefont {X.}~\bibnamefont {Xu}},\ }\href {\doibase
  10.1038/nature22391} {\bibfield  {journal} {\bibinfo  {journal} {Nature}\
  }\textbf {\bibinfo {volume} {546}},\ \bibinfo {pages} {270} (\bibinfo {year}
  {2017})}\BibitemShut {NoStop}%
\bibitem [{\citenamefont {Song}\ \emph {et~al.}(2018)\citenamefont {Song},
  \citenamefont {Cai}, \citenamefont {Tu}, \citenamefont {Zhang}, \citenamefont
  {Huang}, \citenamefont {Wilson}, \citenamefont {Seyler}, \citenamefont {Zhu},
  \citenamefont {Taniguchi}, \citenamefont {Watanabe}, \citenamefont {McGuire},
  \citenamefont {Cobden}, \citenamefont {Xiao}, \citenamefont {Yao},\ and\
  \citenamefont {Xu}}]{song_giant_2018}%
  \BibitemOpen
  \bibfield  {author} {\bibinfo {author} {\bibfnamefont {T.}~\bibnamefont
  {Song}}, \bibinfo {author} {\bibfnamefont {X.}~\bibnamefont {Cai}}, \bibinfo
  {author} {\bibfnamefont {M.~W.-Y.}\ \bibnamefont {Tu}}, \bibinfo {author}
  {\bibfnamefont {X.}~\bibnamefont {Zhang}}, \bibinfo {author} {\bibfnamefont
  {B.}~\bibnamefont {Huang}}, \bibinfo {author} {\bibfnamefont {N.~P.}\
  \bibnamefont {Wilson}}, \bibinfo {author} {\bibfnamefont {K.~L.}\
  \bibnamefont {Seyler}}, \bibinfo {author} {\bibfnamefont {L.}~\bibnamefont
  {Zhu}}, \bibinfo {author} {\bibfnamefont {T.}~\bibnamefont {Taniguchi}},
  \bibinfo {author} {\bibfnamefont {K.}~\bibnamefont {Watanabe}}, \bibinfo
  {author} {\bibfnamefont {M.~A.}\ \bibnamefont {McGuire}}, \bibinfo {author}
  {\bibfnamefont {D.~H.}\ \bibnamefont {Cobden}}, \bibinfo {author}
  {\bibfnamefont {D.}~\bibnamefont {Xiao}}, \bibinfo {author} {\bibfnamefont
  {W.}~\bibnamefont {Yao}}, \ and\ \bibinfo {author} {\bibfnamefont
  {X.}~\bibnamefont {Xu}},\ }\href {\doibase 10.1126/science.aar4851}
  {\bibfield  {journal} {\bibinfo  {journal} {Science}\ }\textbf {\bibinfo
  {volume} {360}},\ \bibinfo {pages} {1214} (\bibinfo {year}
  {2018})}\BibitemShut {NoStop}%
\bibitem [{\citenamefont {Jiang}\ \emph {et~al.}(2018)\citenamefont {Jiang},
  \citenamefont {Shan},\ and\ \citenamefont {Mak}}]{jiang_electric-field_2018}%
  \BibitemOpen
  \bibfield  {author} {\bibinfo {author} {\bibfnamefont {S.}~\bibnamefont
  {Jiang}}, \bibinfo {author} {\bibfnamefont {J.}~\bibnamefont {Shan}}, \ and\
  \bibinfo {author} {\bibfnamefont {K.~F.}\ \bibnamefont {Mak}},\ }\href
  {\doibase 10.1038/s41563-018-0040-6} {\bibfield  {journal} {\bibinfo
  {journal} {Nature Materials}\ }\textbf {\bibinfo {volume} {17}},\ \bibinfo
  {pages} {406} (\bibinfo {year} {2018})}\BibitemShut {NoStop}%
\bibitem [{\citenamefont {Huang}\ \emph {et~al.}(2018)\citenamefont {Huang},
  \citenamefont {Clark}, \citenamefont {Klein}, \citenamefont {MacNeill},
  \citenamefont {Navarro-Moratalla}, \citenamefont {Seyler}, \citenamefont
  {Wilson}, \citenamefont {McGuire}, \citenamefont {Cobden}, \citenamefont
  {Xiao}, \citenamefont {Yao}, \citenamefont {Jarillo-Herrero},\ and\
  \citenamefont {Xu}}]{huang_electrical_2018}%
  \BibitemOpen
  \bibfield  {author} {\bibinfo {author} {\bibfnamefont {B.}~\bibnamefont
  {Huang}}, \bibinfo {author} {\bibfnamefont {G.}~\bibnamefont {Clark}},
  \bibinfo {author} {\bibfnamefont {D.~R.}\ \bibnamefont {Klein}}, \bibinfo
  {author} {\bibfnamefont {D.}~\bibnamefont {MacNeill}}, \bibinfo {author}
  {\bibfnamefont {E.}~\bibnamefont {Navarro-Moratalla}}, \bibinfo {author}
  {\bibfnamefont {K.~L.}\ \bibnamefont {Seyler}}, \bibinfo {author}
  {\bibfnamefont {N.}~\bibnamefont {Wilson}}, \bibinfo {author} {\bibfnamefont
  {M.~A.}\ \bibnamefont {McGuire}}, \bibinfo {author} {\bibfnamefont {D.~H.}\
  \bibnamefont {Cobden}}, \bibinfo {author} {\bibfnamefont {D.}~\bibnamefont
  {Xiao}}, \bibinfo {author} {\bibfnamefont {W.}~\bibnamefont {Yao}}, \bibinfo
  {author} {\bibfnamefont {P.}~\bibnamefont {Jarillo-Herrero}}, \ and\ \bibinfo
  {author} {\bibfnamefont {X.}~\bibnamefont {Xu}},\ }\href {\doibase
  10.1038/s41565-018-0121-3} {\bibfield  {journal} {\bibinfo  {journal} {Nature
  Nanotechnology}\ }\textbf {\bibinfo {volume} {13}},\ \bibinfo {pages} {544}
  (\bibinfo {year} {2018})}\BibitemShut {NoStop}%
\bibitem [{\citenamefont {Gibertini}\ \emph {et~al.}(2019)\citenamefont
  {Gibertini}, \citenamefont {Koperski}, \citenamefont {Morpurgo},\ and\
  \citenamefont {Novoselov}}]{gibertini_magnetic_2019}%
  \BibitemOpen
  \bibfield  {author} {\bibinfo {author} {\bibfnamefont {M.}~\bibnamefont
  {Gibertini}}, \bibinfo {author} {\bibfnamefont {M.}~\bibnamefont {Koperski}},
  \bibinfo {author} {\bibfnamefont {A.~F.}\ \bibnamefont {Morpurgo}}, \ and\
  \bibinfo {author} {\bibfnamefont {K.~S.}\ \bibnamefont {Novoselov}},\ }\href
  {\doibase 10.1038/s41565-019-0438-6} {\bibfield  {journal} {\bibinfo
  {journal} {Nature Nanotechnology}\ }\textbf {\bibinfo {volume} {14}},\
  \bibinfo {pages} {408} (\bibinfo {year} {2019})}\BibitemShut {NoStop}%
\bibitem [{\citenamefont {Chu}\ \emph {et~al.}(2020)\citenamefont {Chu},
  \citenamefont {Roh}, \citenamefont {Island}, \citenamefont {Li},
  \citenamefont {Lee}, \citenamefont {Chen}, \citenamefont {Park},
  \citenamefont {Young}, \citenamefont {Lee},\ and\ \citenamefont
  {Hsieh}}]{chu_linear_2020}%
  \BibitemOpen
  \bibfield  {author} {\bibinfo {author} {\bibfnamefont {H.}~\bibnamefont
  {Chu}}, \bibinfo {author} {\bibfnamefont {C.~J.}\ \bibnamefont {Roh}},
  \bibinfo {author} {\bibfnamefont {J.~O.}\ \bibnamefont {Island}}, \bibinfo
  {author} {\bibfnamefont {C.}~\bibnamefont {Li}}, \bibinfo {author}
  {\bibfnamefont {S.}~\bibnamefont {Lee}}, \bibinfo {author} {\bibfnamefont
  {J.}~\bibnamefont {Chen}}, \bibinfo {author} {\bibfnamefont {J.-G.}\
  \bibnamefont {Park}}, \bibinfo {author} {\bibfnamefont {A.~F.}\ \bibnamefont
  {Young}}, \bibinfo {author} {\bibfnamefont {J.~S.}\ \bibnamefont {Lee}}, \
  and\ \bibinfo {author} {\bibfnamefont {D.}~\bibnamefont {Hsieh}},\ }\href
  {\doibase 10.1103/PhysRevLett.124.027601} {\bibfield  {journal} {\bibinfo
  {journal} {Physical Review Letters}\ }\textbf {\bibinfo {volume} {124}},\
  \bibinfo {pages} {027601} (\bibinfo {year} {2020})}\BibitemShut {NoStop}%
\bibitem [{\citenamefont {Baltz}\ \emph {et~al.}(2018)\citenamefont {Baltz},
  \citenamefont {Manchon}, \citenamefont {Tsoi}, \citenamefont {Moriyama},
  \citenamefont {Ono},\ and\ \citenamefont
  {Tserkovnyak}}]{RMP_Tserkovnyak_2018}%
  \BibitemOpen
  \bibfield  {author} {\bibinfo {author} {\bibfnamefont {V.}~\bibnamefont
  {Baltz}}, \bibinfo {author} {\bibfnamefont {A.}~\bibnamefont {Manchon}},
  \bibinfo {author} {\bibfnamefont {M.}~\bibnamefont {Tsoi}}, \bibinfo {author}
  {\bibfnamefont {T.}~\bibnamefont {Moriyama}}, \bibinfo {author}
  {\bibfnamefont {T.}~\bibnamefont {Ono}}, \ and\ \bibinfo {author}
  {\bibfnamefont {Y.}~\bibnamefont {Tserkovnyak}},\ }\href {\doibase
  10.1103/RevModPhys.90.015005} {\bibfield  {journal} {\bibinfo  {journal}
  {Rev. Mod. Phys.}\ }\textbf {\bibinfo {volume} {90}},\ \bibinfo {pages}
  {015005} (\bibinfo {year} {2018})}\BibitemShut {NoStop}%
\bibitem [{\citenamefont {Jungwirth}\ \emph {et~al.}(2018)\citenamefont
  {Jungwirth}, \citenamefont {Sinova}, \citenamefont {Manchon}, \citenamefont
  {Marti}, \citenamefont {Wunderlich},\ and\ \citenamefont
  {Felser}}]{Jungwirth2018}%
  \BibitemOpen
  \bibfield  {author} {\bibinfo {author} {\bibfnamefont {T.}~\bibnamefont
  {Jungwirth}}, \bibinfo {author} {\bibfnamefont {J.}~\bibnamefont {Sinova}},
  \bibinfo {author} {\bibfnamefont {A.}~\bibnamefont {Manchon}}, \bibinfo
  {author} {\bibfnamefont {X.}~\bibnamefont {Marti}}, \bibinfo {author}
  {\bibfnamefont {J.}~\bibnamefont {Wunderlich}}, \ and\ \bibinfo {author}
  {\bibfnamefont {C.}~\bibnamefont {Felser}},\ }\href {\doibase
  10.1038/s41567-018-0063-6} {\bibfield  {journal} {\bibinfo  {journal} {Nature
  Physics}\ }\textbf {\bibinfo {volume} {14}},\ \bibinfo {pages} {200}
  (\bibinfo {year} {2018})}\BibitemShut {NoStop}%
\bibitem [{\citenamefont {N{\v{e}}mec}\ \emph {et~al.}(2018)\citenamefont
  {N{\v{e}}mec}, \citenamefont {Fiebig}, \citenamefont {Kampfrath},\ and\
  \citenamefont {Kimel}}]{Nemec2018}%
  \BibitemOpen
  \bibfield  {author} {\bibinfo {author} {\bibfnamefont {P.}~\bibnamefont
  {N{\v{e}}mec}}, \bibinfo {author} {\bibfnamefont {M.}~\bibnamefont {Fiebig}},
  \bibinfo {author} {\bibfnamefont {T.}~\bibnamefont {Kampfrath}}, \ and\
  \bibinfo {author} {\bibfnamefont {A.~V.}\ \bibnamefont {Kimel}},\ }\href
  {\doibase 10.1038/s41567-018-0051-x} {\bibfield  {journal} {\bibinfo
  {journal} {Nature Physics}\ }\textbf {\bibinfo {volume} {14}},\ \bibinfo
  {pages} {229} (\bibinfo {year} {2018})}\BibitemShut {NoStop}%
\bibitem [{\citenamefont {Ishizuka}\ and\ \citenamefont
  {Sato}(2019{\natexlab{a}})}]{Ishizuka19b}%
  \BibitemOpen
  \bibfield  {author} {\bibinfo {author} {\bibfnamefont {H.}~\bibnamefont
  {Ishizuka}}\ and\ \bibinfo {author} {\bibfnamefont {M.}~\bibnamefont
  {Sato}},\ }\href {\doibase 10.1103/PhysRevLett.122.197702} {\bibfield
  {journal} {\bibinfo  {journal} {Phys. Rev. Lett.}\ }\textbf {\bibinfo
  {volume} {122}},\ \bibinfo {pages} {197702} (\bibinfo {year}
  {2019}{\natexlab{a}})}\BibitemShut {NoStop}%
\bibitem [{\citenamefont {McIver}\ \emph {et~al.}(2012)\citenamefont {McIver},
  \citenamefont {Hsieh}, \citenamefont {Steinberg}, \citenamefont
  {Jarillo-Herrero},\ and\ \citenamefont {Gedik}}]{mciver_control_2012}%
  \BibitemOpen
  \bibfield  {author} {\bibinfo {author} {\bibfnamefont {J.~W.}\ \bibnamefont
  {McIver}}, \bibinfo {author} {\bibfnamefont {D.}~\bibnamefont {Hsieh}},
  \bibinfo {author} {\bibfnamefont {H.}~\bibnamefont {Steinberg}}, \bibinfo
  {author} {\bibfnamefont {P.}~\bibnamefont {Jarillo-Herrero}}, \ and\ \bibinfo
  {author} {\bibfnamefont {N.}~\bibnamefont {Gedik}},\ }\href {\doibase
  10.1038/nnano.2011.214} {\bibfield  {journal} {\bibinfo  {journal} {Nature
  Nanotechnology}\ }\textbf {\bibinfo {volume} {7}},\ \bibinfo {pages} {96}
  (\bibinfo {year} {2012})}\BibitemShut {NoStop}%
\bibitem [{\citenamefont {Xu}\ \emph {et~al.}(2018)\citenamefont {Xu},
  \citenamefont {Ma}, \citenamefont {Shen}, \citenamefont {Fatemi},
  \citenamefont {Wu}, \citenamefont {Chang}, \citenamefont {Chang},
  \citenamefont {Valdivia}, \citenamefont {Chan}, \citenamefont {Gibson},
  \citenamefont {Zhou}, \citenamefont {Liu}, \citenamefont {Watanabe},
  \citenamefont {Taniguchi}, \citenamefont {Lin}, \citenamefont {Cava},
  \citenamefont {Fu}, \citenamefont {Gedik},\ and\ \citenamefont
  {Jarillo-Herrero}}]{xu_electrically_2018}%
  \BibitemOpen
  \bibfield  {author} {\bibinfo {author} {\bibfnamefont {S.-Y.}\ \bibnamefont
  {Xu}}, \bibinfo {author} {\bibfnamefont {Q.}~\bibnamefont {Ma}}, \bibinfo
  {author} {\bibfnamefont {H.}~\bibnamefont {Shen}}, \bibinfo {author}
  {\bibfnamefont {V.}~\bibnamefont {Fatemi}}, \bibinfo {author} {\bibfnamefont
  {S.}~\bibnamefont {Wu}}, \bibinfo {author} {\bibfnamefont {T.-R.}\
  \bibnamefont {Chang}}, \bibinfo {author} {\bibfnamefont {G.}~\bibnamefont
  {Chang}}, \bibinfo {author} {\bibfnamefont {A.~M.~M.}\ \bibnamefont
  {Valdivia}}, \bibinfo {author} {\bibfnamefont {C.-K.}\ \bibnamefont {Chan}},
  \bibinfo {author} {\bibfnamefont {Q.~D.}\ \bibnamefont {Gibson}}, \bibinfo
  {author} {\bibfnamefont {J.}~\bibnamefont {Zhou}}, \bibinfo {author}
  {\bibfnamefont {Z.}~\bibnamefont {Liu}}, \bibinfo {author} {\bibfnamefont
  {K.}~\bibnamefont {Watanabe}}, \bibinfo {author} {\bibfnamefont
  {T.}~\bibnamefont {Taniguchi}}, \bibinfo {author} {\bibfnamefont
  {H.}~\bibnamefont {Lin}}, \bibinfo {author} {\bibfnamefont {R.~J.}\
  \bibnamefont {Cava}}, \bibinfo {author} {\bibfnamefont {L.}~\bibnamefont
  {Fu}}, \bibinfo {author} {\bibfnamefont {N.}~\bibnamefont {Gedik}}, \ and\
  \bibinfo {author} {\bibfnamefont {P.}~\bibnamefont {Jarillo-Herrero}},\
  }\href {\doibase 10.1038/s41567-018-0189-6} {\bibfield  {journal} {\bibinfo
  {journal} {Nature Physics}\ }\textbf {\bibinfo {volume} {14}},\ \bibinfo
  {pages} {900} (\bibinfo {year} {2018})}\BibitemShut {NoStop}%
\bibitem [{\citenamefont {Ma}\ \emph {et~al.}(2017)\citenamefont {Ma},
  \citenamefont {Xu}, \citenamefont {Chan}, \citenamefont {Zhang},
  \citenamefont {Chang}, \citenamefont {Lin}, \citenamefont {Xie},
  \citenamefont {Palacios}, \citenamefont {Lin}, \citenamefont {Jia},
  \citenamefont {Lee}, \citenamefont {Jarillo-Herrero},\ and\ \citenamefont
  {Gedik}}]{ma_direct_2017}%
  \BibitemOpen
  \bibfield  {author} {\bibinfo {author} {\bibfnamefont {Q.}~\bibnamefont
  {Ma}}, \bibinfo {author} {\bibfnamefont {S.-Y.}\ \bibnamefont {Xu}}, \bibinfo
  {author} {\bibfnamefont {C.-K.}\ \bibnamefont {Chan}}, \bibinfo {author}
  {\bibfnamefont {C.-L.}\ \bibnamefont {Zhang}}, \bibinfo {author}
  {\bibfnamefont {G.}~\bibnamefont {Chang}}, \bibinfo {author} {\bibfnamefont
  {Y.}~\bibnamefont {Lin}}, \bibinfo {author} {\bibfnamefont {W.}~\bibnamefont
  {Xie}}, \bibinfo {author} {\bibfnamefont {T.}~\bibnamefont {Palacios}},
  \bibinfo {author} {\bibfnamefont {H.}~\bibnamefont {Lin}}, \bibinfo {author}
  {\bibfnamefont {S.}~\bibnamefont {Jia}}, \bibinfo {author} {\bibfnamefont
  {P.~A.}\ \bibnamefont {Lee}}, \bibinfo {author} {\bibfnamefont
  {P.}~\bibnamefont {Jarillo-Herrero}}, \ and\ \bibinfo {author} {\bibfnamefont
  {N.}~\bibnamefont {Gedik}},\ }\href {\doibase 10.1038/nphys4146} {\bibfield
  {journal} {\bibinfo  {journal} {Nature Physics}\ }\textbf {\bibinfo {volume}
  {13}},\ \bibinfo {pages} {842} (\bibinfo {year} {2017})}\BibitemShut
  {NoStop}%
\bibitem [{\citenamefont {Kirilyuk}\ \emph {et~al.}(2010)\citenamefont
  {Kirilyuk}, \citenamefont {Kimel},\ and\ \citenamefont
  {Rasing}}]{kirilyuk_ultrafast_2010}%
  \BibitemOpen
  \bibfield  {author} {\bibinfo {author} {\bibfnamefont {A.}~\bibnamefont
  {Kirilyuk}}, \bibinfo {author} {\bibfnamefont {A.~V.}\ \bibnamefont {Kimel}},
  \ and\ \bibinfo {author} {\bibfnamefont {T.}~\bibnamefont {Rasing}},\ }\href
  {\doibase 10.1103/RevModPhys.82.2731} {\bibfield  {journal} {\bibinfo
  {journal} {Reviews of Modern Physics}\ }\textbf {\bibinfo {volume} {82}},\
  \bibinfo {pages} {2731} (\bibinfo {year} {2010})}\BibitemShut {NoStop}%
\bibitem [{\citenamefont {Nova}\ \emph {et~al.}(2017)\citenamefont {Nova},
  \citenamefont {Cartella}, \citenamefont {Cantaluppi}, \citenamefont
  {F{\"o}rst}, \citenamefont {Bossini}, \citenamefont {Mikhaylovskiy},
  \citenamefont {Kimel}, \citenamefont {Merlin},\ and\ \citenamefont
  {Cavalleri}}]{nova_effective_2017}%
  \BibitemOpen
  \bibfield  {author} {\bibinfo {author} {\bibfnamefont {T.~F.}\ \bibnamefont
  {Nova}}, \bibinfo {author} {\bibfnamefont {A.}~\bibnamefont {Cartella}},
  \bibinfo {author} {\bibfnamefont {A.}~\bibnamefont {Cantaluppi}}, \bibinfo
  {author} {\bibfnamefont {M.}~\bibnamefont {F{\"o}rst}}, \bibinfo {author}
  {\bibfnamefont {D.}~\bibnamefont {Bossini}}, \bibinfo {author} {\bibfnamefont
  {R.~V.}\ \bibnamefont {Mikhaylovskiy}}, \bibinfo {author} {\bibfnamefont
  {A.~V.}\ \bibnamefont {Kimel}}, \bibinfo {author} {\bibfnamefont
  {R.}~\bibnamefont {Merlin}}, \ and\ \bibinfo {author} {\bibfnamefont
  {A.}~\bibnamefont {Cavalleri}},\ }\href {\doibase 10.1038/nphys3925}
  {\bibfield  {journal} {\bibinfo  {journal} {Nat. Phys.}\ }\textbf {\bibinfo
  {volume} {13}},\ \bibinfo {pages} {132} (\bibinfo {year} {2017})}\BibitemShut
  {NoStop}%
\bibitem [{\citenamefont {Afanasiev}\ \emph {et~al.}(2021)\citenamefont
  {Afanasiev}, \citenamefont {Hortensius}, \citenamefont {Ivanov},
  \citenamefont {Sasani}, \citenamefont {Bousquet}, \citenamefont {Blanter},
  \citenamefont {Mikhaylovskiy}, \citenamefont {Kimel},\ and\ \citenamefont
  {Caviglia}}]{afanasiev2019lightdriven}%
  \BibitemOpen
  \bibfield  {author} {\bibinfo {author} {\bibfnamefont {D.}~\bibnamefont
  {Afanasiev}}, \bibinfo {author} {\bibfnamefont {J.~R.}\ \bibnamefont
  {Hortensius}}, \bibinfo {author} {\bibfnamefont {B.~A.}\ \bibnamefont
  {Ivanov}}, \bibinfo {author} {\bibfnamefont {A.}~\bibnamefont {Sasani}},
  \bibinfo {author} {\bibfnamefont {E.}~\bibnamefont {Bousquet}}, \bibinfo
  {author} {\bibfnamefont {Y.~M.}\ \bibnamefont {Blanter}}, \bibinfo {author}
  {\bibfnamefont {R.~V.}\ \bibnamefont {Mikhaylovskiy}}, \bibinfo {author}
  {\bibfnamefont {A.~V.}\ \bibnamefont {Kimel}}, \ and\ \bibinfo {author}
  {\bibfnamefont {A.~D.}\ \bibnamefont {Caviglia}},\ }\href {\doibase
  10.1038/s41563-021-00922-7} {\bibfield  {journal} {\bibinfo  {journal} {Nat.
  Mater.}\ } (\bibinfo {year} {2021}),\ 10.1038/s41563-021-00922-7}\BibitemShut
  {NoStop}%
\bibitem [{\citenamefont {Disa}\ \emph {et~al.}(2020)\citenamefont {Disa},
  \citenamefont {Fechner}, \citenamefont {Nova}, \citenamefont {Liu},
  \citenamefont {F{\"o}rst}, \citenamefont {Prabhakaran}, \citenamefont
  {Radaelli},\ and\ \citenamefont {Cavalleri}}]{disa_polarizing_2020}%
  \BibitemOpen
  \bibfield  {author} {\bibinfo {author} {\bibfnamefont {A.~S.}\ \bibnamefont
  {Disa}}, \bibinfo {author} {\bibfnamefont {M.}~\bibnamefont {Fechner}},
  \bibinfo {author} {\bibfnamefont {T.~F.}\ \bibnamefont {Nova}}, \bibinfo
  {author} {\bibfnamefont {B.}~\bibnamefont {Liu}}, \bibinfo {author}
  {\bibfnamefont {M.}~\bibnamefont {F{\"o}rst}}, \bibinfo {author}
  {\bibfnamefont {D.}~\bibnamefont {Prabhakaran}}, \bibinfo {author}
  {\bibfnamefont {P.~G.}\ \bibnamefont {Radaelli}}, \ and\ \bibinfo {author}
  {\bibfnamefont {A.}~\bibnamefont {Cavalleri}},\ }\href {\doibase
  10.1038/s41567-020-0936-3} {\bibfield  {journal} {\bibinfo  {journal} {Nat.
  Phys.}\ }\textbf {\bibinfo {volume} {16}},\ \bibinfo {pages} {937} (\bibinfo
  {year} {2020})}\BibitemShut {NoStop}%
\bibitem [{\citenamefont {Stupakiewicz}\ \emph {et~al.}(2021)\citenamefont
  {Stupakiewicz}, \citenamefont {Davies}, \citenamefont {Szerenos},
  \citenamefont {Afanasiev}, \citenamefont {Rabinovich}, \citenamefont {Boris},
  \citenamefont {Caviglia}, \citenamefont {Kimel},\ and\ \citenamefont
  {Kirilyuk}}]{stupakiewicz_ultrafast_2021}%
  \BibitemOpen
  \bibfield  {author} {\bibinfo {author} {\bibfnamefont {A.}~\bibnamefont
  {Stupakiewicz}}, \bibinfo {author} {\bibfnamefont {C.~S.}\ \bibnamefont
  {Davies}}, \bibinfo {author} {\bibfnamefont {K.}~\bibnamefont {Szerenos}},
  \bibinfo {author} {\bibfnamefont {D.}~\bibnamefont {Afanasiev}}, \bibinfo
  {author} {\bibfnamefont {K.~S.}\ \bibnamefont {Rabinovich}}, \bibinfo
  {author} {\bibfnamefont {A.~V.}\ \bibnamefont {Boris}}, \bibinfo {author}
  {\bibfnamefont {A.}~\bibnamefont {Caviglia}}, \bibinfo {author}
  {\bibfnamefont {A.~V.}\ \bibnamefont {Kimel}}, \ and\ \bibinfo {author}
  {\bibfnamefont {A.}~\bibnamefont {Kirilyuk}},\ }\href {\doibase
  10.1038/s41567-020-01124-9} {\bibfield  {journal} {\bibinfo  {journal} {Nat.
  Phys.}\ } (\bibinfo {year} {2021}),\ 10.1038/s41567-020-01124-9}\BibitemShut
  {NoStop}%
\bibitem [{\citenamefont {Walowski}\ and\ \citenamefont
  {Münzenberg}(2016)}]{Walowski2016}%
  \BibitemOpen
  \bibfield  {author} {\bibinfo {author} {\bibfnamefont {J.}~\bibnamefont
  {Walowski}}\ and\ \bibinfo {author} {\bibfnamefont {M.}~\bibnamefont
  {Münzenberg}},\ }\href {\doibase 10.1063/1.4958846} {\bibfield  {journal}
  {\bibinfo  {journal} {J. Appl. Phys.}\ }\textbf {\bibinfo {volume} {120}},\
  \bibinfo {pages} {140901} (\bibinfo {year} {2016})}\BibitemShut {NoStop}%
\bibitem [{\citenamefont {Schlauderer}\ \emph {et~al.}(2019)\citenamefont
  {Schlauderer}, \citenamefont {Lange}, \citenamefont {Baierl}, \citenamefont
  {Ebnet}, \citenamefont {Schmid}, \citenamefont {Valovcin}, \citenamefont
  {Zvezdin}, \citenamefont {Kimel}, \citenamefont {Mikhaylovskiy},\ and\
  \citenamefont {Huber}}]{Schlauderer2019}%
  \BibitemOpen
  \bibfield  {author} {\bibinfo {author} {\bibfnamefont {S.}~\bibnamefont
  {Schlauderer}}, \bibinfo {author} {\bibfnamefont {C.}~\bibnamefont {Lange}},
  \bibinfo {author} {\bibfnamefont {S.}~\bibnamefont {Baierl}}, \bibinfo
  {author} {\bibfnamefont {T.}~\bibnamefont {Ebnet}}, \bibinfo {author}
  {\bibfnamefont {C.~P.}\ \bibnamefont {Schmid}}, \bibinfo {author}
  {\bibfnamefont {D.~C.}\ \bibnamefont {Valovcin}}, \bibinfo {author}
  {\bibfnamefont {A.~K.}\ \bibnamefont {Zvezdin}}, \bibinfo {author}
  {\bibfnamefont {A.~V.}\ \bibnamefont {Kimel}}, \bibinfo {author}
  {\bibfnamefont {R.~V.}\ \bibnamefont {Mikhaylovskiy}}, \ and\ \bibinfo
  {author} {\bibfnamefont {R.}~\bibnamefont {Huber}},\ }\href {\doibase
  10.1038/s41586-019-1174-7} {\bibfield  {journal} {\bibinfo  {journal}
  {Nature}\ }\textbf {\bibinfo {volume} {569}},\ \bibinfo {pages} {383}
  (\bibinfo {year} {2019})}\BibitemShut {NoStop}%
\bibitem [{\citenamefont {Siegrist}\ \emph {et~al.}(2019)\citenamefont
  {Siegrist}, \citenamefont {Gessner}, \citenamefont {Ossiander}, \citenamefont
  {Denker}, \citenamefont {Chang}, \citenamefont {Schr{\"o}der}, \citenamefont
  {Guggenmos}, \citenamefont {Cui}, \citenamefont {Walowski}, \citenamefont
  {Martens}, \citenamefont {Dewhurst}, \citenamefont {Kleineberg},
  \citenamefont {M{\"u}nzenberg}, \citenamefont {Sharma},\ and\ \citenamefont
  {Schultze}}]{Siegrist2019}%
  \BibitemOpen
  \bibfield  {author} {\bibinfo {author} {\bibfnamefont {F.}~\bibnamefont
  {Siegrist}}, \bibinfo {author} {\bibfnamefont {J.~A.}\ \bibnamefont
  {Gessner}}, \bibinfo {author} {\bibfnamefont {M.}~\bibnamefont {Ossiander}},
  \bibinfo {author} {\bibfnamefont {C.}~\bibnamefont {Denker}}, \bibinfo
  {author} {\bibfnamefont {Y.-P.}\ \bibnamefont {Chang}}, \bibinfo {author}
  {\bibfnamefont {M.~C.}\ \bibnamefont {Schr{\"o}der}}, \bibinfo {author}
  {\bibfnamefont {A.}~\bibnamefont {Guggenmos}}, \bibinfo {author}
  {\bibfnamefont {Y.}~\bibnamefont {Cui}}, \bibinfo {author} {\bibfnamefont
  {J.}~\bibnamefont {Walowski}}, \bibinfo {author} {\bibfnamefont
  {U.}~\bibnamefont {Martens}}, \bibinfo {author} {\bibfnamefont {J.~K.}\
  \bibnamefont {Dewhurst}}, \bibinfo {author} {\bibfnamefont {U.}~\bibnamefont
  {Kleineberg}}, \bibinfo {author} {\bibfnamefont {M.}~\bibnamefont
  {M{\"u}nzenberg}}, \bibinfo {author} {\bibfnamefont {S.}~\bibnamefont
  {Sharma}}, \ and\ \bibinfo {author} {\bibfnamefont {M.}~\bibnamefont
  {Schultze}},\ }\href {\doibase 10.1038/s41586-019-1333-x} {\bibfield
  {journal} {\bibinfo  {journal} {Nature}\ }\textbf {\bibinfo {volume} {571}},\
  \bibinfo {pages} {240} (\bibinfo {year} {2019})}\BibitemShut {NoStop}%
\bibitem [{\citenamefont {Fleury}\ and\ \citenamefont
  {Loudon}(1968)}]{Fleury68}%
  \BibitemOpen
  \bibfield  {author} {\bibinfo {author} {\bibfnamefont {P.~A.}\ \bibnamefont
  {Fleury}}\ and\ \bibinfo {author} {\bibfnamefont {R.}~\bibnamefont
  {Loudon}},\ }\href {\doibase 10.1103/PhysRev.166.514} {\bibfield  {journal}
  {\bibinfo  {journal} {Phys. Rev.}\ }\textbf {\bibinfo {volume} {166}},\
  \bibinfo {pages} {514} (\bibinfo {year} {1968})}\BibitemShut {NoStop}%
\bibitem [{\citenamefont {Proskurin}\ \emph {et~al.}(2018)\citenamefont
  {Proskurin}, \citenamefont {Ovchinnikov}, \citenamefont {Kishine},\ and\
  \citenamefont {Stamps}}]{Proskurin18}%
  \BibitemOpen
  \bibfield  {author} {\bibinfo {author} {\bibfnamefont {I.}~\bibnamefont
  {Proskurin}}, \bibinfo {author} {\bibfnamefont {A.~S.}\ \bibnamefont
  {Ovchinnikov}}, \bibinfo {author} {\bibfnamefont {J.-i.}\ \bibnamefont
  {Kishine}}, \ and\ \bibinfo {author} {\bibfnamefont {R.~L.}\ \bibnamefont
  {Stamps}},\ }\href {\doibase 10.1103/PhysRevB.98.134422} {\bibfield
  {journal} {\bibinfo  {journal} {Phys. Rev. B}\ }\textbf {\bibinfo {volume}
  {98}},\ \bibinfo {pages} {134422} (\bibinfo {year} {2018})}\BibitemShut
  {NoStop}%
\bibitem [{\citenamefont {Ishizuka}\ and\ \citenamefont
  {Sato}(2019{\natexlab{b}})}]{Ishizuka19}%
  \BibitemOpen
  \bibfield  {author} {\bibinfo {author} {\bibfnamefont {H.}~\bibnamefont
  {Ishizuka}}\ and\ \bibinfo {author} {\bibfnamefont {M.}~\bibnamefont
  {Sato}},\ }\href {\doibase 10.1103/PhysRevB.100.224411} {\bibfield  {journal}
  {\bibinfo  {journal} {Phys. Rev. B}\ }\textbf {\bibinfo {volume} {100}},\
  \bibinfo {pages} {224411} (\bibinfo {year} {2019}{\natexlab{b}})}\BibitemShut
  {NoStop}%
\bibitem [{\citenamefont {Proskurin}\ and\ \citenamefont
  {Stamps}(2020)}]{proskurin2020symmetry}%
  \BibitemOpen
  \bibfield  {author} {\bibinfo {author} {\bibfnamefont {I.}~\bibnamefont
  {Proskurin}}\ and\ \bibinfo {author} {\bibfnamefont {R.~L.}\ \bibnamefont
  {Stamps}},\ }\href@noop {} {\bibfield  {journal} {\bibinfo  {journal} {arXiv
  preprint arXiv:2006.07399}\ } (\bibinfo {year} {2020})}\BibitemShut {NoStop}%
\bibitem [{\citenamefont {Fleury}\ \emph {et~al.}(1967)\citenamefont {Fleury},
  \citenamefont {Porto},\ and\ \citenamefont {Loudon}}]{Fleury67}%
  \BibitemOpen
  \bibfield  {author} {\bibinfo {author} {\bibfnamefont {P.~A.}\ \bibnamefont
  {Fleury}}, \bibinfo {author} {\bibfnamefont {S.~P.~S.}\ \bibnamefont
  {Porto}}, \ and\ \bibinfo {author} {\bibfnamefont {R.}~\bibnamefont
  {Loudon}},\ }\href {\doibase 10.1103/PhysRevLett.18.658} {\bibfield
  {journal} {\bibinfo  {journal} {Phys. Rev. Lett.}\ }\textbf {\bibinfo
  {volume} {18}},\ \bibinfo {pages} {658} (\bibinfo {year} {1967})}\BibitemShut
  {NoStop}%
\bibitem [{\citenamefont {Shastry}\ and\ \citenamefont
  {Shraiman}(1990)}]{Shastry90}%
  \BibitemOpen
  \bibfield  {author} {\bibinfo {author} {\bibfnamefont {B.~S.}\ \bibnamefont
  {Shastry}}\ and\ \bibinfo {author} {\bibfnamefont {B.~I.}\ \bibnamefont
  {Shraiman}},\ }\href {\doibase 10.1103/PhysRevLett.65.1068} {\bibfield
  {journal} {\bibinfo  {journal} {Phys. Rev. Lett.}\ }\textbf {\bibinfo
  {volume} {65}},\ \bibinfo {pages} {1068} (\bibinfo {year}
  {1990})}\BibitemShut {NoStop}%
\bibitem [{\citenamefont {Fei}\ \emph {et~al.}(2021)\citenamefont {Fei},
  \citenamefont {Song}, \citenamefont {Pusey-Nazzaro},\ and\ \citenamefont
  {Yang}}]{Fei21}%
  \BibitemOpen
  \bibfield  {author} {\bibinfo {author} {\bibfnamefont {R.}~\bibnamefont
  {Fei}}, \bibinfo {author} {\bibfnamefont {W.}~\bibnamefont {Song}}, \bibinfo
  {author} {\bibfnamefont {L.}~\bibnamefont {Pusey-Nazzaro}}, \ and\ \bibinfo
  {author} {\bibfnamefont {L.}~\bibnamefont {Yang}},\ }\href@noop {} {\enquote
  {\bibinfo {title} {Pt-symmetry enabled spin circular photogalvanic effect in
  antiferromagnetic insulators},}\ } (\bibinfo {year} {2021}),\ \Eprint
  {http://arxiv.org/abs/arXiv:2104.08341} {arXiv:2104.08341} \BibitemShut
  {NoStop}%
\bibitem [{SM()}]{SM}%
  \BibitemOpen
  \href@noop {} {}\Eprint {http://arxiv.org/abs/Supplemental Material at [URL
  will be inserted by publisher] for [give brief description of material].}
  {Supplemental Material at [URL will be inserted by publisher] for [give brief
  description of material].} \BibitemShut {NoStop}%
\bibitem [{\citenamefont {Cornelissen}\ \emph {et~al.}(2015)\citenamefont
  {Cornelissen}, \citenamefont {Liu}, \citenamefont {Duine}, \citenamefont
  {Youssef},\ and\ \citenamefont {van Wees}}]{Cornelissen15}%
  \BibitemOpen
  \bibfield  {author} {\bibinfo {author} {\bibfnamefont {L.~J.}\ \bibnamefont
  {Cornelissen}}, \bibinfo {author} {\bibfnamefont {J.}~\bibnamefont {Liu}},
  \bibinfo {author} {\bibfnamefont {R.~A.}\ \bibnamefont {Duine}}, \bibinfo
  {author} {\bibfnamefont {J.~B.}\ \bibnamefont {Youssef}}, \ and\ \bibinfo
  {author} {\bibfnamefont {B.~J.}\ \bibnamefont {van Wees}},\ }\href {\doibase
  10.1038/nphys3465} {\bibfield  {journal} {\bibinfo  {journal} {Nature
  Physics}\ }\textbf {\bibinfo {volume} {11}},\ \bibinfo {pages} {1022}
  (\bibinfo {year} {2015})}\BibitemShut {NoStop}%
\bibitem [{\citenamefont {Lebrun}\ \emph {et~al.}(2018)\citenamefont {Lebrun},
  \citenamefont {Ross}, \citenamefont {Bender}, \citenamefont {Qaiumzadeh},
  \citenamefont {Baldrati}, \citenamefont {Cramer}, \citenamefont {Brataas},
  \citenamefont {Duine},\ and\ \citenamefont {Kl\"{a}ui}}]{Lebrun18}%
  \BibitemOpen
  \bibfield  {author} {\bibinfo {author} {\bibfnamefont {R.}~\bibnamefont
  {Lebrun}}, \bibinfo {author} {\bibfnamefont {A.}~\bibnamefont {Ross}},
  \bibinfo {author} {\bibfnamefont {S.~A.}\ \bibnamefont {Bender}}, \bibinfo
  {author} {\bibfnamefont {A.}~\bibnamefont {Qaiumzadeh}}, \bibinfo {author}
  {\bibfnamefont {L.}~\bibnamefont {Baldrati}}, \bibinfo {author}
  {\bibfnamefont {J.}~\bibnamefont {Cramer}}, \bibinfo {author} {\bibfnamefont
  {A.}~\bibnamefont {Brataas}}, \bibinfo {author} {\bibfnamefont {R.~A.}\
  \bibnamefont {Duine}}, \ and\ \bibinfo {author} {\bibfnamefont
  {M.}~\bibnamefont {Kl\"{a}ui}},\ }\href {\doibase 10.1038/s41586-018-0490-7}
  {\bibfield  {journal} {\bibinfo  {journal} {Nature}\ }\textbf {\bibinfo
  {volume} {561}},\ \bibinfo {pages} {222} (\bibinfo {year}
  {2018})}\BibitemShut {NoStop}%
\bibitem [{\citenamefont {Ando}\ \emph {et~al.}(2011)\citenamefont {Ando},
  \citenamefont {Takahashi}, \citenamefont {Ieda}, \citenamefont {Kajiwara},
  \citenamefont {Nakayama}, \citenamefont {Yoshino}, \citenamefont {Harii},
  \citenamefont {Fujikawa}, \citenamefont {Matsuo}, \citenamefont {Maekawa},\
  and\ \citenamefont {Saitoh}}]{Ando11}%
  \BibitemOpen
  \bibfield  {author} {\bibinfo {author} {\bibfnamefont {K.}~\bibnamefont
  {Ando}}, \bibinfo {author} {\bibfnamefont {S.}~\bibnamefont {Takahashi}},
  \bibinfo {author} {\bibfnamefont {J.}~\bibnamefont {Ieda}}, \bibinfo {author}
  {\bibfnamefont {Y.}~\bibnamefont {Kajiwara}}, \bibinfo {author}
  {\bibfnamefont {H.}~\bibnamefont {Nakayama}}, \bibinfo {author}
  {\bibfnamefont {T.}~\bibnamefont {Yoshino}}, \bibinfo {author} {\bibfnamefont
  {K.}~\bibnamefont {Harii}}, \bibinfo {author} {\bibfnamefont
  {Y.}~\bibnamefont {Fujikawa}}, \bibinfo {author} {\bibfnamefont
  {M.}~\bibnamefont {Matsuo}}, \bibinfo {author} {\bibfnamefont
  {S.}~\bibnamefont {Maekawa}}, \ and\ \bibinfo {author} {\bibfnamefont
  {E.}~\bibnamefont {Saitoh}},\ }\href {\doibase 10.1063/1.3587173} {\bibfield
  {journal} {\bibinfo  {journal} {Journal of Applied Physics}\ }\textbf
  {\bibinfo {volume} {109}},\ \bibinfo {pages} {103913} (\bibinfo {year}
  {2011})}\BibitemShut {NoStop}%
\bibitem [{\citenamefont {Zhang}\ and\ \citenamefont {Zhang}(2012)}]{Zhang12}%
  \BibitemOpen
  \bibfield  {author} {\bibinfo {author} {\bibfnamefont {S.~S.-L.}\
  \bibnamefont {Zhang}}\ and\ \bibinfo {author} {\bibfnamefont
  {S.}~\bibnamefont {Zhang}},\ }\href {\doibase 10.1103/physrevb.86.214424}
  {\bibfield  {journal} {\bibinfo  {journal} {Physical Review B}\ }\textbf
  {\bibinfo {volume} {86}},\ \bibinfo {pages} {214424} (\bibinfo {year}
  {2012})}\BibitemShut {NoStop}%
\bibitem [{\citenamefont {Wei}\ \emph {et~al.}(2014)\citenamefont {Wei},
  \citenamefont {Obstbaum}, \citenamefont {Ribow}, \citenamefont {Back},\ and\
  \citenamefont {Woltersdorf}}]{Wei14}%
  \BibitemOpen
  \bibfield  {author} {\bibinfo {author} {\bibfnamefont {D.}~\bibnamefont
  {Wei}}, \bibinfo {author} {\bibfnamefont {M.}~\bibnamefont {Obstbaum}},
  \bibinfo {author} {\bibfnamefont {M.}~\bibnamefont {Ribow}}, \bibinfo
  {author} {\bibfnamefont {C.~H.}\ \bibnamefont {Back}}, \ and\ \bibinfo
  {author} {\bibfnamefont {G.}~\bibnamefont {Woltersdorf}},\ }\href {\doibase
  10.1038/ncomms4768} {\bibfield  {journal} {\bibinfo  {journal} {Nature
  Communications}\ }\textbf {\bibinfo {volume} {5}} (\bibinfo {year} {2014}),\
  10.1038/ncomms4768}\BibitemShut {NoStop}%
\bibitem [{\citenamefont {Xing}\ \emph {et~al.}(2019)\citenamefont {Xing},
  \citenamefont {Qiu}, \citenamefont {Wang}, \citenamefont {Yao}, \citenamefont
  {Ma}, \citenamefont {Cai}, \citenamefont {Jia}, \citenamefont {Xie},\ and\
  \citenamefont {Han}}]{Xing19}%
  \BibitemOpen
  \bibfield  {author} {\bibinfo {author} {\bibfnamefont {W.}~\bibnamefont
  {Xing}}, \bibinfo {author} {\bibfnamefont {L.}~\bibnamefont {Qiu}}, \bibinfo
  {author} {\bibfnamefont {X.}~\bibnamefont {Wang}}, \bibinfo {author}
  {\bibfnamefont {Y.}~\bibnamefont {Yao}}, \bibinfo {author} {\bibfnamefont
  {Y.}~\bibnamefont {Ma}}, \bibinfo {author} {\bibfnamefont {R.}~\bibnamefont
  {Cai}}, \bibinfo {author} {\bibfnamefont {S.}~\bibnamefont {Jia}}, \bibinfo
  {author} {\bibfnamefont {X.~C.}\ \bibnamefont {Xie}}, \ and\ \bibinfo
  {author} {\bibfnamefont {W.}~\bibnamefont {Han}},\ }\href {\doibase
  10.1103/PhysRevX.9.011026} {\bibfield  {journal} {\bibinfo  {journal} {Phys.
  Rev. X}\ }\textbf {\bibinfo {volume} {9}},\ \bibinfo {pages} {011026}
  (\bibinfo {year} {2019})}\BibitemShut {NoStop}%
\bibitem [{\citenamefont {Wildes}\ \emph
  {et~al.}(1998{\natexlab{a}})\citenamefont {Wildes}, \citenamefont {Roessli},
  \citenamefont {Lebech},\ and\ \citenamefont {Godfrey}}]{Wildes98}%
  \BibitemOpen
  \bibfield  {author} {\bibinfo {author} {\bibfnamefont {A.~R.}\ \bibnamefont
  {Wildes}}, \bibinfo {author} {\bibfnamefont {B.}~\bibnamefont {Roessli}},
  \bibinfo {author} {\bibfnamefont {B.}~\bibnamefont {Lebech}}, \ and\ \bibinfo
  {author} {\bibfnamefont {K.~W.}\ \bibnamefont {Godfrey}},\ }\href {\doibase
  10.1088/0953-8984/10/28/020} {\bibfield  {journal} {\bibinfo  {journal}
  {Journal of Physics: Condensed Matter}\ }\textbf {\bibinfo {volume} {10}},\
  \bibinfo {pages} {6417} (\bibinfo {year} {1998}{\natexlab{a}})}\BibitemShut
  {NoStop}%
\bibitem [{\citenamefont {Cheng}\ \emph {et~al.}(2016)\citenamefont {Cheng},
  \citenamefont {Okamoto},\ and\ \citenamefont {Xiao}}]{Cheng16}%
  \BibitemOpen
  \bibfield  {author} {\bibinfo {author} {\bibfnamefont {R.}~\bibnamefont
  {Cheng}}, \bibinfo {author} {\bibfnamefont {S.}~\bibnamefont {Okamoto}}, \
  and\ \bibinfo {author} {\bibfnamefont {D.}~\bibnamefont {Xiao}},\ }\href
  {\doibase 10.1103/PhysRevLett.117.217202} {\bibfield  {journal} {\bibinfo
  {journal} {Phys. Rev. Lett.}\ }\textbf {\bibinfo {volume} {117}},\ \bibinfo
  {pages} {217202} (\bibinfo {year} {2016})}\BibitemShut {NoStop}%
\bibitem [{\citenamefont {Wildes}\ \emph
  {et~al.}(1998{\natexlab{b}})\citenamefont {Wildes}, \citenamefont {Roessli},
  \citenamefont {Lebech},\ and\ \citenamefont {Godfrey}}]{Wildes}%
  \BibitemOpen
  \bibfield  {author} {\bibinfo {author} {\bibfnamefont {A.~R.}\ \bibnamefont
  {Wildes}}, \bibinfo {author} {\bibfnamefont {B.}~\bibnamefont {Roessli}},
  \bibinfo {author} {\bibfnamefont {B.}~\bibnamefont {Lebech}}, \ and\ \bibinfo
  {author} {\bibfnamefont {K.~W.}\ \bibnamefont {Godfrey}},\ }\href {\doibase
  10.1088/0953-8984/10/28/020} {\bibfield  {journal} {\bibinfo  {journal}
  {Journal of Physics: Condensed Matter}\ }\textbf {\bibinfo {volume} {10}},\
  \bibinfo {pages} {6417} (\bibinfo {year} {1998}{\natexlab{b}})}\BibitemShut
  {NoStop}%
\bibitem [{\citenamefont {Kim}\ \emph {et~al.}(2019)\citenamefont {Kim},
  \citenamefont {Lim}, \citenamefont {Kim}, \citenamefont {Lee}, \citenamefont
  {Lee}, \citenamefont {Kim}, \citenamefont {Park}, \citenamefont {Son},
  \citenamefont {Park}, \citenamefont {Park},\ and\ \citenamefont
  {Cheong}}]{Kim_2019}%
  \BibitemOpen
  \bibfield  {author} {\bibinfo {author} {\bibfnamefont {K.}~\bibnamefont
  {Kim}}, \bibinfo {author} {\bibfnamefont {S.~Y.}\ \bibnamefont {Lim}},
  \bibinfo {author} {\bibfnamefont {J.}~\bibnamefont {Kim}}, \bibinfo {author}
  {\bibfnamefont {J.-U.}\ \bibnamefont {Lee}}, \bibinfo {author} {\bibfnamefont
  {S.}~\bibnamefont {Lee}}, \bibinfo {author} {\bibfnamefont {P.}~\bibnamefont
  {Kim}}, \bibinfo {author} {\bibfnamefont {K.}~\bibnamefont {Park}}, \bibinfo
  {author} {\bibfnamefont {S.}~\bibnamefont {Son}}, \bibinfo {author}
  {\bibfnamefont {C.-H.}\ \bibnamefont {Park}}, \bibinfo {author}
  {\bibfnamefont {J.-G.}\ \bibnamefont {Park}}, \ and\ \bibinfo {author}
  {\bibfnamefont {H.}~\bibnamefont {Cheong}},\ }\href {\doibase
  10.1088/2053-1583/ab27d5} {\bibfield  {journal} {\bibinfo  {journal} {2D
  Materials}\ }\textbf {\bibinfo {volume} {6}},\ \bibinfo {pages} {041001}
  (\bibinfo {year} {2019})}\BibitemShut {NoStop}%
\bibitem [{\citenamefont {Chittari}\ \emph {et~al.}(2016)\citenamefont
  {Chittari}, \citenamefont {Park}, \citenamefont {Lee}, \citenamefont {Han},
  \citenamefont {MacDonald}, \citenamefont {Hwang},\ and\ \citenamefont
  {Jung}}]{chittari2016electronic}%
  \BibitemOpen
  \bibfield  {author} {\bibinfo {author} {\bibfnamefont {B.~L.}\ \bibnamefont
  {Chittari}}, \bibinfo {author} {\bibfnamefont {Y.}~\bibnamefont {Park}},
  \bibinfo {author} {\bibfnamefont {D.}~\bibnamefont {Lee}}, \bibinfo {author}
  {\bibfnamefont {M.}~\bibnamefont {Han}}, \bibinfo {author} {\bibfnamefont
  {A.~H.}\ \bibnamefont {MacDonald}}, \bibinfo {author} {\bibfnamefont
  {E.}~\bibnamefont {Hwang}}, \ and\ \bibinfo {author} {\bibfnamefont
  {J.}~\bibnamefont {Jung}},\ }\href {\doibase 10.1103/PhysRevB.94.184428}
  {\bibfield  {journal} {\bibinfo  {journal} {Phys. Rev. B}\ }\textbf {\bibinfo
  {volume} {94}},\ \bibinfo {pages} {184428} (\bibinfo {year}
  {2016})}\BibitemShut {NoStop}%
\bibitem [{\citenamefont {Bazazzadeh}\ \emph {et~al.}(2021)\citenamefont
  {Bazazzadeh}, \citenamefont {Hamdi}, \citenamefont {Haddadi}, \citenamefont
  {Khavasi}, \citenamefont {Sadeghi},\ and\ \citenamefont
  {Mohseni}}]{bazazzadeh2021symmetry}%
  \BibitemOpen
  \bibfield  {author} {\bibinfo {author} {\bibfnamefont {N.}~\bibnamefont
  {Bazazzadeh}}, \bibinfo {author} {\bibfnamefont {M.}~\bibnamefont {Hamdi}},
  \bibinfo {author} {\bibfnamefont {F.}~\bibnamefont {Haddadi}}, \bibinfo
  {author} {\bibfnamefont {A.}~\bibnamefont {Khavasi}}, \bibinfo {author}
  {\bibfnamefont {A.}~\bibnamefont {Sadeghi}}, \ and\ \bibinfo {author}
  {\bibfnamefont {S.~M.}\ \bibnamefont {Mohseni}},\ }\href {\doibase
  10.1103/PhysRevB.103.014425} {\bibfield  {journal} {\bibinfo  {journal}
  {Phys. Rev. B}\ }\textbf {\bibinfo {volume} {103}},\ \bibinfo {pages}
  {014425} (\bibinfo {year} {2021})}\BibitemShut {NoStop}%
\bibitem [{Note1()}]{Note1}%
  \BibitemOpen
  \bibinfo {note} {This follows from the fact that in honeycomb
  antiferromagnets, the terms of the Hamiltonian describing the DMI are
  proportional to the identity matrix.}\BibitemShut {Stop}%
\bibitem [{\citenamefont {Yao}\ \emph {et~al.}(2008)\citenamefont {Yao},
  \citenamefont {Xiao},\ and\ \citenamefont {Niu}}]{Yao08}%
  \BibitemOpen
  \bibfield  {author} {\bibinfo {author} {\bibfnamefont {W.}~\bibnamefont
  {Yao}}, \bibinfo {author} {\bibfnamefont {D.}~\bibnamefont {Xiao}}, \ and\
  \bibinfo {author} {\bibfnamefont {Q.}~\bibnamefont {Niu}},\ }\href {\doibase
  10.1103/PhysRevB.77.235406} {\bibfield  {journal} {\bibinfo  {journal} {Phys.
  Rev. B}\ }\textbf {\bibinfo {volume} {77}},\ \bibinfo {pages} {235406}
  (\bibinfo {year} {2008})}\BibitemShut {NoStop}%
\bibitem [{\citenamefont {Topp}\ \emph {et~al.}(2021)\citenamefont {Topp},
  \citenamefont {Eckhardt}, \citenamefont {Kennes}, \citenamefont {Sentef},\
  and\ \citenamefont {T{\"o}rm{\"a}}}]{Topp21}%
  \BibitemOpen
  \bibfield  {author} {\bibinfo {author} {\bibfnamefont {G.~E.}\ \bibnamefont
  {Topp}}, \bibinfo {author} {\bibfnamefont {C.~J.}\ \bibnamefont {Eckhardt}},
  \bibinfo {author} {\bibfnamefont {D.~M.}\ \bibnamefont {Kennes}}, \bibinfo
  {author} {\bibfnamefont {M.~A.}\ \bibnamefont {Sentef}}, \ and\ \bibinfo
  {author} {\bibfnamefont {P.}~\bibnamefont {T{\"o}rm{\"a}}},\ }\href@noop {}
  {\enquote {\bibinfo {title} {Light-matter coupling and quantum geometry in
  moir{\'e} materials},}\ } (\bibinfo {year} {2021}),\ \Eprint
  {http://arxiv.org/abs/arXiv:2103.04967} {arXiv:2103.04967} \BibitemShut
  {NoStop}%
\bibitem [{\citenamefont {Li}\ \emph {et~al.}(2020)\citenamefont {Li},
  \citenamefont {Tohyama}, \citenamefont {Iitaka}, \citenamefont {Su},\ and\
  \citenamefont {Zeng}}]{li_detection_2020}%
  \BibitemOpen
  \bibfield  {author} {\bibinfo {author} {\bibfnamefont {Z.}~\bibnamefont
  {Li}}, \bibinfo {author} {\bibfnamefont {T.}~\bibnamefont {Tohyama}},
  \bibinfo {author} {\bibfnamefont {T.}~\bibnamefont {Iitaka}}, \bibinfo
  {author} {\bibfnamefont {H.}~\bibnamefont {Su}}, \ and\ \bibinfo {author}
  {\bibfnamefont {H.}~\bibnamefont {Zeng}},\ }\href
  {http://arxiv.org/abs/2007.02481} {\bibfield  {journal} {\bibinfo  {journal}
  {arXiv:2007.02481 [cond-mat, physics:physics]}\ } (\bibinfo {year}
  {2020})}\BibitemShut {NoStop}%
\bibitem [{\citenamefont {Ahn}\ \emph {et~al.}(2020)\citenamefont {Ahn},
  \citenamefont {Guo},\ and\ \citenamefont {Nagaosa}}]{ahn_low-frequency_2020}%
  \BibitemOpen
  \bibfield  {author} {\bibinfo {author} {\bibfnamefont {J.}~\bibnamefont
  {Ahn}}, \bibinfo {author} {\bibfnamefont {G.-Y.}\ \bibnamefont {Guo}}, \ and\
  \bibinfo {author} {\bibfnamefont {N.}~\bibnamefont {Nagaosa}},\ }\href
  {\doibase 10.1103/PhysRevX.10.041041} {\bibfield  {journal} {\bibinfo
  {journal} {Physical Review X}\ }\textbf {\bibinfo {volume} {10}},\ \bibinfo
  {pages} {041041} (\bibinfo {year} {2020})}\BibitemShut {NoStop}%
\bibitem [{\citenamefont {Ahn}\ \emph {et~al.}(2021)\citenamefont {Ahn},
  \citenamefont {Guo}, \citenamefont {Nagaosa},\ and\ \citenamefont
  {Vishwanath}}]{ahn_riemannian_2021}%
  \BibitemOpen
  \bibfield  {author} {\bibinfo {author} {\bibfnamefont {J.}~\bibnamefont
  {Ahn}}, \bibinfo {author} {\bibfnamefont {G.-Y.}\ \bibnamefont {Guo}},
  \bibinfo {author} {\bibfnamefont {N.}~\bibnamefont {Nagaosa}}, \ and\
  \bibinfo {author} {\bibfnamefont {A.}~\bibnamefont {Vishwanath}},\ }\href
  {http://arxiv.org/abs/2103.01241} {\bibfield  {journal} {\bibinfo  {journal}
  {arXiv:2103.01241 [cond-mat, physics:physics]}\ } (\bibinfo {year} {2021})},\
  \bibinfo {note} {arXiv: 2103.01241}\BibitemShut {NoStop}%
\bibitem [{\citenamefont {{Vi{\~n}as Bostr{\"o}m}}\ \emph
  {et~al.}(2020)\citenamefont {{Vi{\~n}as Bostr{\"o}m}}, \citenamefont
  {Claassen}, \citenamefont {McIver}, \citenamefont {Jotzu}, \citenamefont
  {Rubio},\ and\ \citenamefont {Sentef}}]{Bostrom20}%
  \BibitemOpen
  \bibfield  {author} {\bibinfo {author} {\bibfnamefont {E.}~\bibnamefont
  {{Vi{\~n}as Bostr{\"o}m}}}, \bibinfo {author} {\bibfnamefont
  {M.}~\bibnamefont {Claassen}}, \bibinfo {author} {\bibfnamefont {J.~W.}\
  \bibnamefont {McIver}}, \bibinfo {author} {\bibfnamefont {G.}~\bibnamefont
  {Jotzu}}, \bibinfo {author} {\bibfnamefont {A.}~\bibnamefont {Rubio}}, \ and\
  \bibinfo {author} {\bibfnamefont {M.~A.}\ \bibnamefont {Sentef}},\ }\href
  {\doibase 10.21468/SciPostPhys.9.4.061} {\bibfield  {journal} {\bibinfo
  {journal} {SciPost Phys.}\ }\textbf {\bibinfo {volume} {9}},\ \bibinfo
  {pages} {61} (\bibinfo {year} {2020})}\BibitemShut {NoStop}%
\bibitem [{\citenamefont {de~Juan}\ \emph {et~al.}(2017)\citenamefont
  {de~Juan}, \citenamefont {Grushin}, \citenamefont {Morimoto},\ and\
  \citenamefont {Moore}}]{de_juan_quantized_2017}%
  \BibitemOpen
  \bibfield  {author} {\bibinfo {author} {\bibfnamefont {F.}~\bibnamefont
  {de~Juan}}, \bibinfo {author} {\bibfnamefont {A.~G.}\ \bibnamefont
  {Grushin}}, \bibinfo {author} {\bibfnamefont {T.}~\bibnamefont {Morimoto}}, \
  and\ \bibinfo {author} {\bibfnamefont {J.~E.}\ \bibnamefont {Moore}},\ }\href
  {\doibase 10.1038/ncomms15995} {\bibfield  {journal} {\bibinfo  {journal}
  {Nature Communications}\ }\textbf {\bibinfo {volume} {8}},\ \bibinfo {pages}
  {15995} (\bibinfo {year} {2017})}\BibitemShut {NoStop}%
\bibitem [{\citenamefont {Ni}\ \emph {et~al.}(2021)\citenamefont {Ni},
  \citenamefont {Wang}, \citenamefont {Zhang}, \citenamefont {Pozo},
  \citenamefont {Xu}, \citenamefont {Han}, \citenamefont {Manna}, \citenamefont
  {Paglione}, \citenamefont {Felser}, \citenamefont {Grushin}, \citenamefont
  {de~Juan}, \citenamefont {Mele},\ and\ \citenamefont {Wu}}]{ni_giant_2021}%
  \BibitemOpen
  \bibfield  {author} {\bibinfo {author} {\bibfnamefont {Z.}~\bibnamefont
  {Ni}}, \bibinfo {author} {\bibfnamefont {K.}~\bibnamefont {Wang}}, \bibinfo
  {author} {\bibfnamefont {Y.}~\bibnamefont {Zhang}}, \bibinfo {author}
  {\bibfnamefont {O.}~\bibnamefont {Pozo}}, \bibinfo {author} {\bibfnamefont
  {B.}~\bibnamefont {Xu}}, \bibinfo {author} {\bibfnamefont {X.}~\bibnamefont
  {Han}}, \bibinfo {author} {\bibfnamefont {K.}~\bibnamefont {Manna}}, \bibinfo
  {author} {\bibfnamefont {J.}~\bibnamefont {Paglione}}, \bibinfo {author}
  {\bibfnamefont {C.}~\bibnamefont {Felser}}, \bibinfo {author} {\bibfnamefont
  {A.~G.}\ \bibnamefont {Grushin}}, \bibinfo {author} {\bibfnamefont
  {F.}~\bibnamefont {de~Juan}}, \bibinfo {author} {\bibfnamefont {E.~J.}\
  \bibnamefont {Mele}}, \ and\ \bibinfo {author} {\bibfnamefont
  {L.}~\bibnamefont {Wu}},\ }\href {\doibase 10.1038/s41467-020-20408-5}
  {\bibfield  {journal} {\bibinfo  {journal} {Nature Communications}\ }\textbf
  {\bibinfo {volume} {12}},\ \bibinfo {pages} {154} (\bibinfo {year}
  {2021})}\BibitemShut {NoStop}%
\end{thebibliography}%

\end{document}